\documentclass[aps,pre,twocolumn,groupedaddress,showpacs,floatfix]{revtex4}
\usepackage{amsmath}
\usepackage{amssymb}
\usepackage{graphicx}

\begin{document}

\title{Growth-Driven Percolations: The Dynamics of Community
Formation in Neuronal Systems}

\author{Luciano da Fontoura Costa} 
\affiliation{Institute of Physics of S\~ao Carlos. 
University of S\~ ao Paulo, S\~{a}o Carlos,
SP, PO Box 369, 13560-970, 
phone +55 16 3373 9858,FAX +55 16 3371
3616, Brazil, luciano@if.sc.usp.br}

\author{Regina C\'elia Coelho}
\affiliation{ Faculdade de Ci\^encias Matem\'aticas, 
da Natureza e Tecnologia de Informa\c{c}\~ao - 
Methodist University of Piracicaba, 
Rodovia do A\c{c}\'ucar, km 156 - Campus Taquaral  - 
Taquaral, Piracicaba, SP, Brazil, 
13400 - 901 phone +55 19 3124 1515 - Ext. 1245,     Fax +55 19 3124 1701}

\date{\today}

\begin{abstract}   

The quintessential property of neuronal systems is their intensive
patterns of selective synaptic connections.  The current work
describes a physics-based approach to neuronal shape modeling and
synthesis and its consideration for the simulation of neuronal
development and the formation of neuronal communities.  Starting from
images of real neurons, geometrical measurements are obtained and used
to construct probabilistic models which can be subsequently sampled in
order to produce morphologically realistic neuronal cells.  Such cells
are progressively grown while monitoring their connections along time,
which are analysed in terms of percolation concepts.  However, unlike
traditional percolation, the critical point is verified along the
growth stages, not the density of cells, which remains constant
throughout the neuronal growth dynamics. It is shown, through
simulations, that growing beta cells tend to reach percolation sooner
than the alpha counterparts with the same diameter.  Also, the percolation
becomes more abrupt for higher densities of cells, being markedly
sharper for the beta cells.

\end{abstract}


\maketitle

\begin{verse}
The brain is a world consisting of a number of unexplored continents
and great stretches of unknown territory.
 \emph{(Santiago Ramon-y-Cajal)}
\end{verse}

Neurons can be understood as cells which, along the evolutionary
process, have become highly specialized for establishing connections
between themselves along a wide range of spatial scales (ranging from
microns to meters).  In order to minimize metabolism and allow
connections to selective targets, neurons acquired their intricated,
ramified shapes.  Indeed, instead of implementing casual connections
with every surrounding cell, a neuron links to specific targets which
can be nearby in the same neuronal region or far away in another
cortical area originating, in the process, the basic architecture
required for proper operation of the central nervous system.
Interestingly, the connectivity pattern of a mature neuronal system is
determined not only by the genome, which is unable to code all
connections
\footnote{Indeed, even the fate of neuroblasts along the cell
differentiation process often does not involve genetical coding
\cite{Kandel:1995}.}, but predominantly by the history of neuronal
activity under stimuli presentation.  Neurons are produced at
ventricular zones of the neuroepithelium, in the form of
\emph{neuroblasts}, which therefore differentiate and migrate to
specific target regions and start to unfold their dendritic and axonal
processes \cite{Kandel:1995}.  As such structures develop and extend
towards specific targets, which occurs under the guidance of trophic
factors, they synapse and start forming communities (or clusters) of
connected cells, organized in specific ways so as to achieve proper
operation.  Indeed, the functional properties of such structures are
to a large extent related to the underlying connecting patterns,
implying that one of the fundamental problems in neuroscience is to
understand how neuronal connections are established during the
development of the central nervous system \cite{Kandel:1995}.

Since connectivity is the main purpose underlying neuronal growth and
organization, it is interesting to obtain suitable mathematical
structures and relationships capable of representing and modeling the
development of neuronal systems at a high level of morphological
realism.  While graphs/networks, where neurons are assigned to nodes
and synapses to edges, provide a natural means to express the neuronal
connections, several concepts from statistical mechanics can be used
to model and simulate the connection dynamics.  By providing an
interesting interface between graph theory and statistical mechanics,
the recent area of complex networks \cite{Albert_Barab:2002,
Newman:2003, Dorog_Mendes:2002} represents a particularly promising
perspective to bridge the gap between the morphology and dynamics of
neuronal systems.  In particular, the concept of percolation
\cite{Book_Stauffer} stand out as particularly relevant for such
investigations.  Previous related works include the statistical
physics investigation of scaling properties and the degree of
separation in cortical networks~\cite{Karbowski:2001}, the small-world
characterization of neuronal structures grown \emph{in vitro}
\cite{Shefi:2002}, the use of critical percolation point for neuronal
shape characterization \cite{Percolation:2003}, and the identification
of electrically active clusters in neural networks \cite{Segev:2003}.
Related works addressing the relationship between neuronal geometry
and function can also be found in the literature
(e.g. \cite{Stauffer_Costa:2003}, \cite{Costa_Stauffer:2003},
\cite{Costa_BM:2003},\cite{Ascoli_Krichmar:2000}).  While such works
have considered stactic neuronal shapes, the development of a
framework to model neuromorphically realistic neurons reported in
\cite{Coelho_Costa:2001} allows investigations of the neuronal
connectivity during simulated neuronal development, by monitoring the
size and other properties of the existing clusters in terms of time.
Such a perspective motivated the extension of the concept of
percolation to consider growing structures where the shapes of the
objects may vary with time, a possibility proposed and investigated
possibly for the first time in the present work.

\begin{figure}
 \begin{center} \includegraphics[scale=.6]{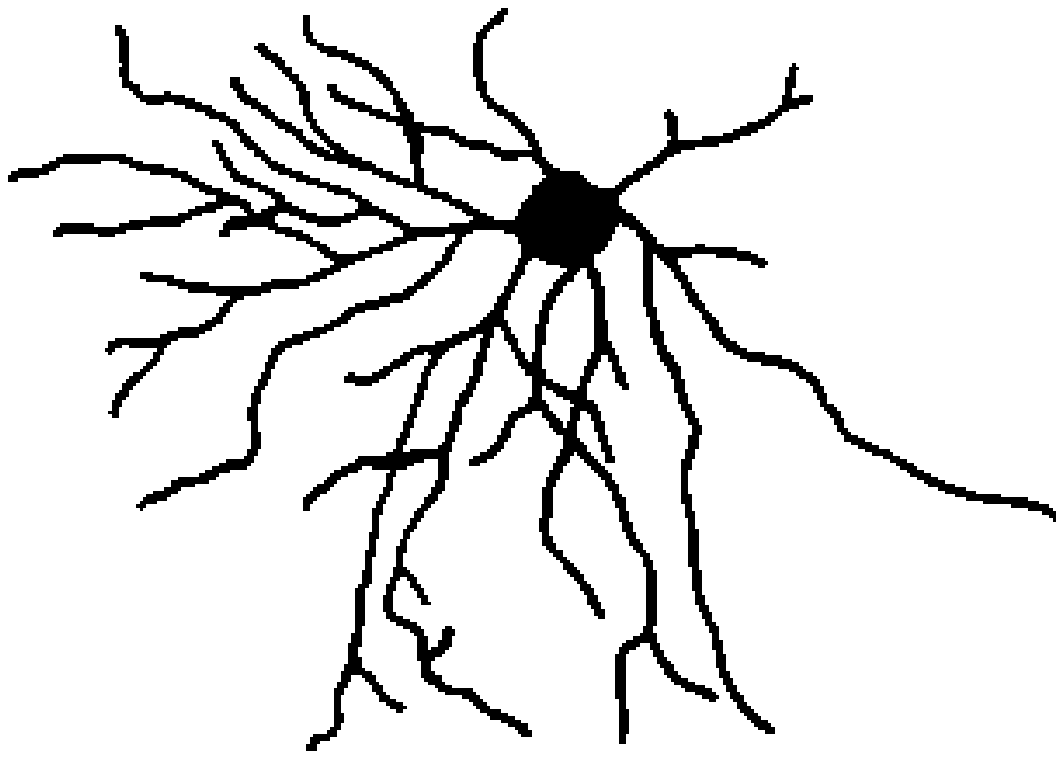}\\
   (a) \\
   \vspace{0.5cm}
   \includegraphics[scale=3]{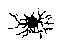} \\
   (b)

   \caption{Example of alpha (a), \cite{Wasslea:1981}, and beta (b)
   \cite{Wassleb:1981} cells used in this work.  Reproduced with
   permission.~\label{fig:ex_cells}}
\end{center}
\end{figure}

This article starts by describing how the neuronal cells are
represented and statistically modeled in terms of probabilities and
follows by presenting the simulation of neuronal growth by using the
Monte Carlo approach, as well as the characterization of the obtained
structures in terms of the maximum cluster size observed along time.
Such issues are illustrated with respect to a database of 2D neuronal
cells including cat retina ganglionar cells of the types alpha (23
samples) and beta (27 samples), of which typical cells are illustrated
in Figure~\ref{fig:ex_cells}.

\section{Neuronal Modeling}

\begin{figure}
 \begin{center} 
   \includegraphics[scale=.25]{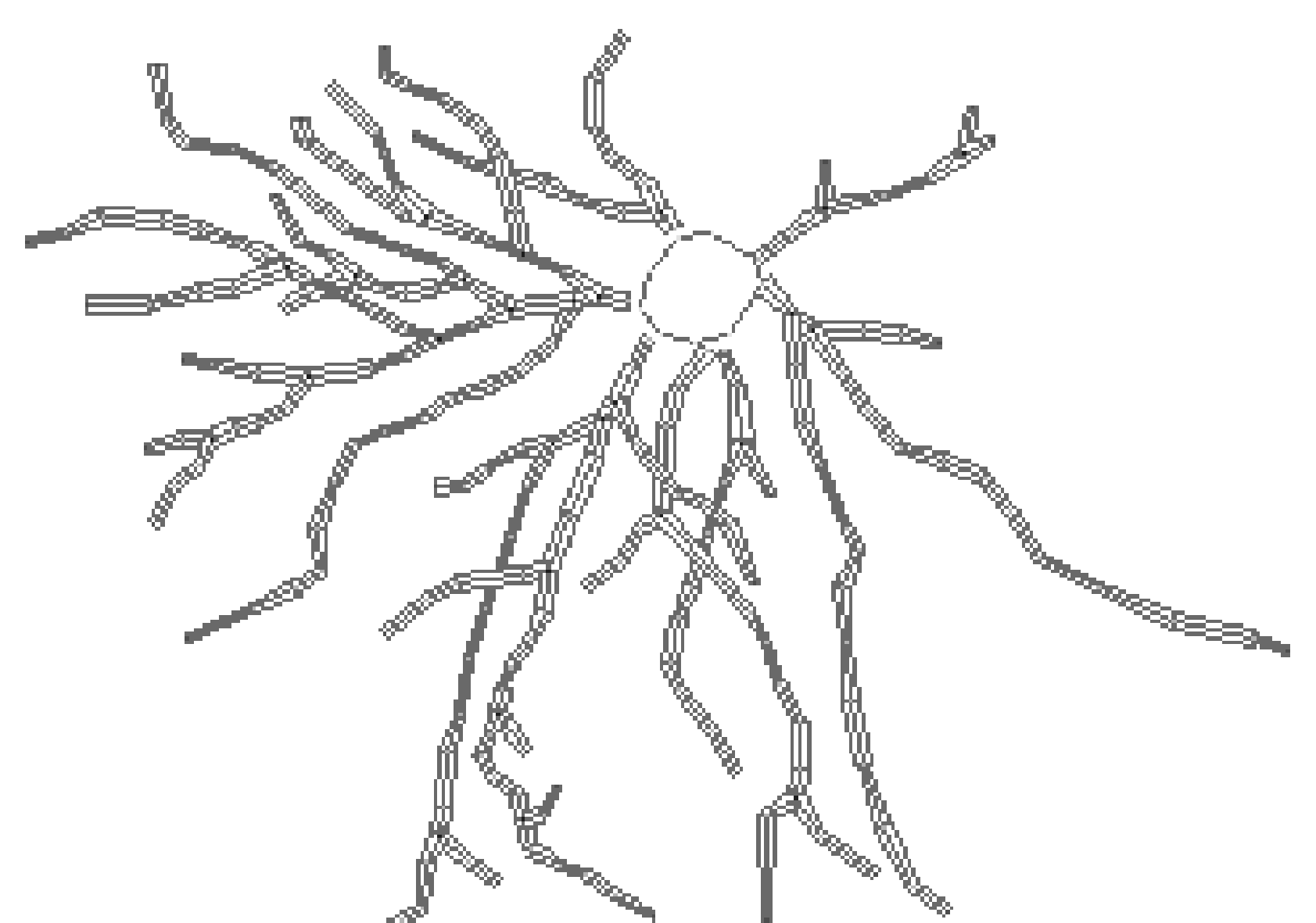} \\
   (a) \\
   \vspace{0.5cm}
   \includegraphics[scale=.12]{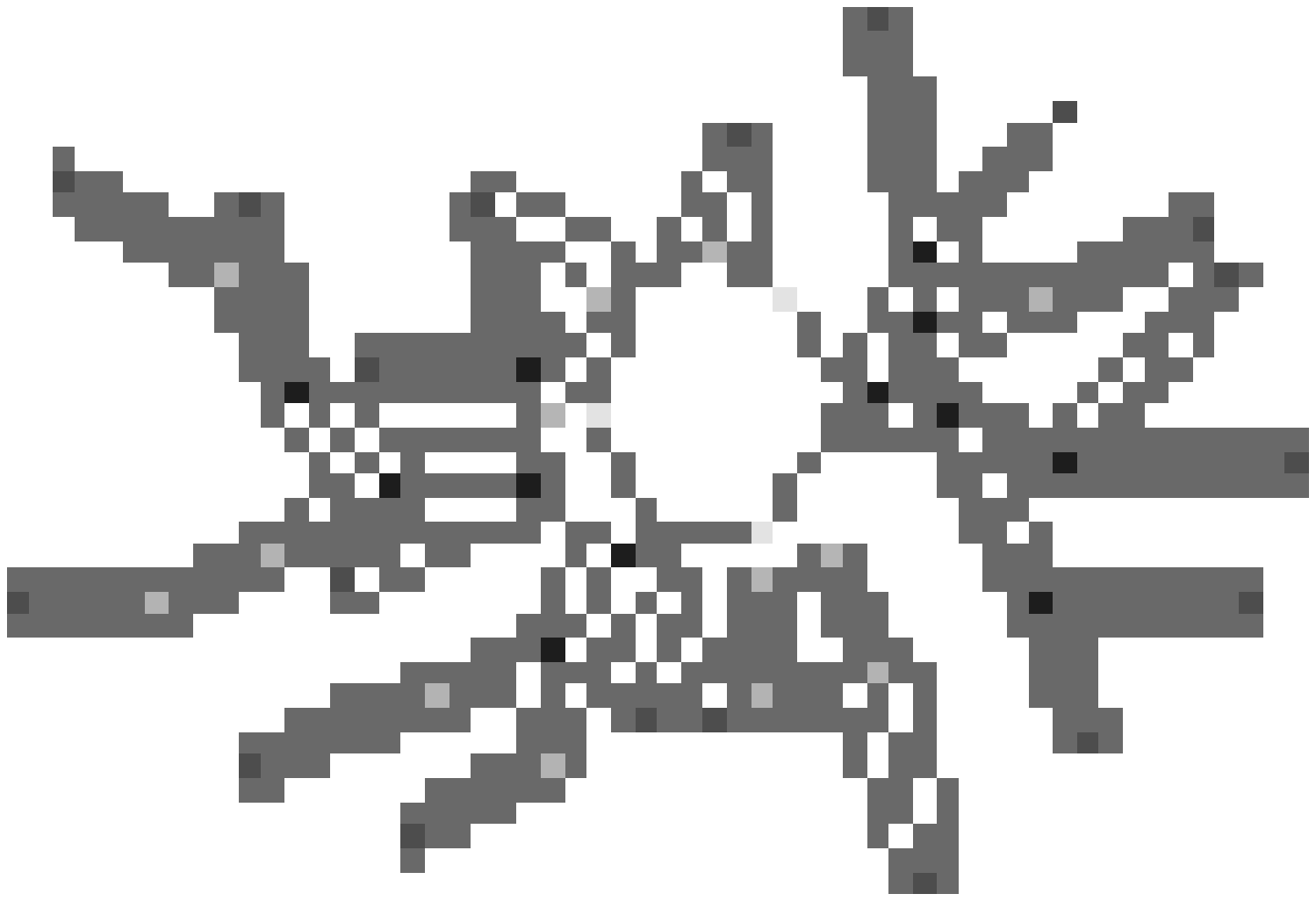} \\
   (b)

   \caption{Traced cells obtained from the cells in
   Figure~\ref{fig:ex_cells} by using a customized neural
   tracer.~\label{fig:ex_impro}} \end{center}
\end{figure}

One first key isse in neuromorphic modeling regards how to represent
mathematically the geometry of neurons.  While the typically observed
diversity of shapes for the same class of cells immediately implies
the use of statistics, the choice of the best (in the sense of being
the most compact) set of morphometric parameters capable of
representing the neuronal shape without considerable loss of
information remains a challenging issue \cite{Ascoli_Krichmar:2000}.
The methodology for 2D neuronal representation adopted in this work
follows the framework reported in \cite{Coelho_Costa:2001}, involving
a probabilistic model considering the number of branches, the angles
between them, the length of the dendritic and axonal segments,
branching probability, and the length and angle of arcs of each
branch.  Therefore, the first step is to obtain such measurements from
images of the real neuronal cells to be modeled.  Typically, the cells
are histologically marked and prepared, mounted on slides, and the
respective images acquired through a camera interfaced to a light
transmission (or fluorescence) microscope.  The neurons in such images
are then identified and isolated (e.g.  \cite{CostaCesar:2001}),
producing binary representations (i.e.  images containing only the
neuronal cell -- marked as one, and the background -- marked as zero).
An alternative way to obtain the binary images of the neural cells is
through camera-lucida drawings, as is the case for the images in
Figure~\ref{fig:ex_cells}.  Once such binary images are obtained,
their boundaries are extracted by using a customized neural
tracer~\footnote{This neural tracer, developed by L.  A. Consularo
during his PhD at the Cybernetic Vision Research group, allows
operator-assisted tracing of the neuronal cells, producing results in
the Eutectic format.}, producing results such as those illustrated in
Figure~\ref{fig:ex_impro}, which corresponds to the boundaries of the
cells in Figure~\ref{fig:ex_cells}.

\begin{figure}
 \begin{center} 
   \includegraphics[scale=.4]{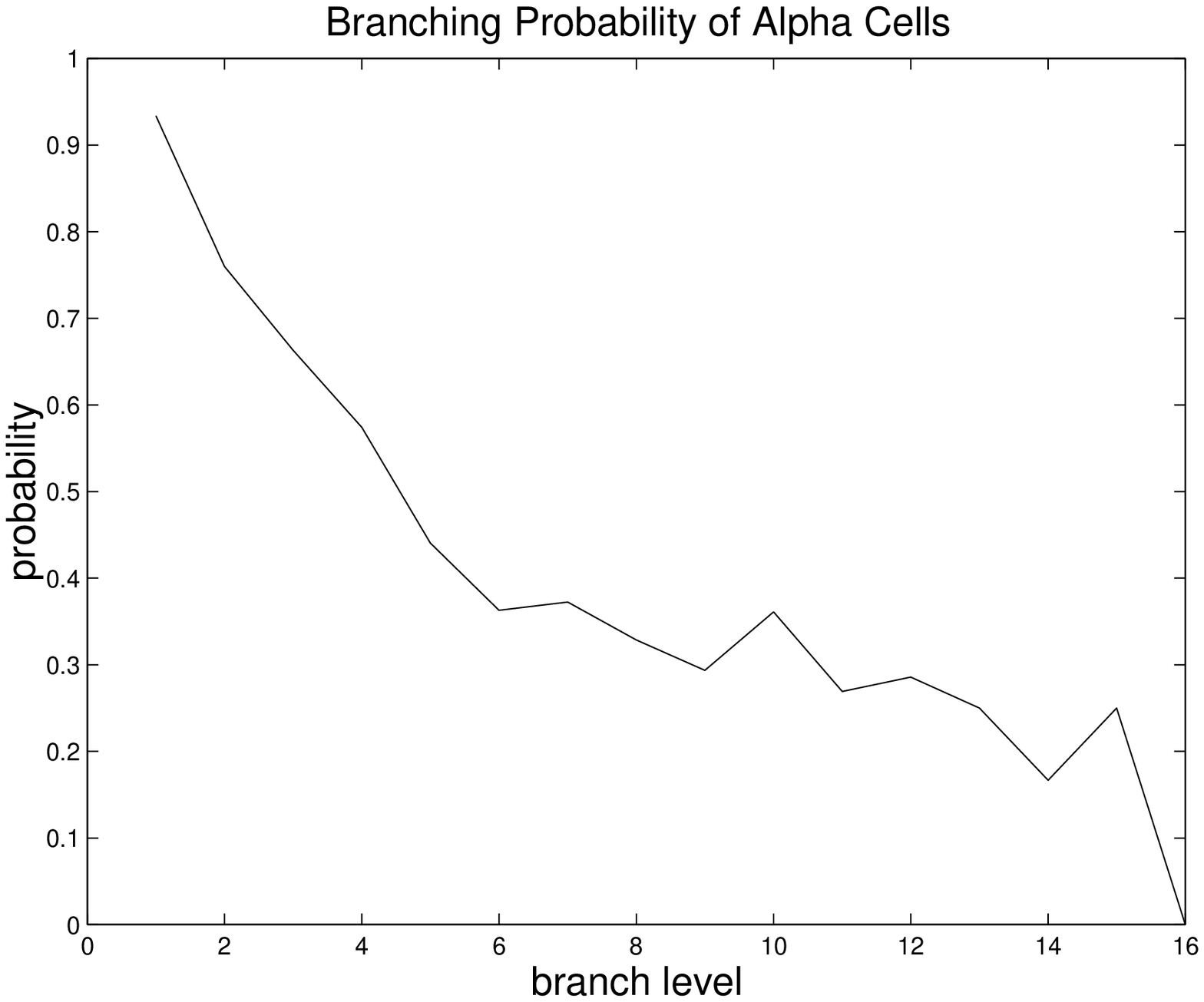}
   \includegraphics[scale=.4]{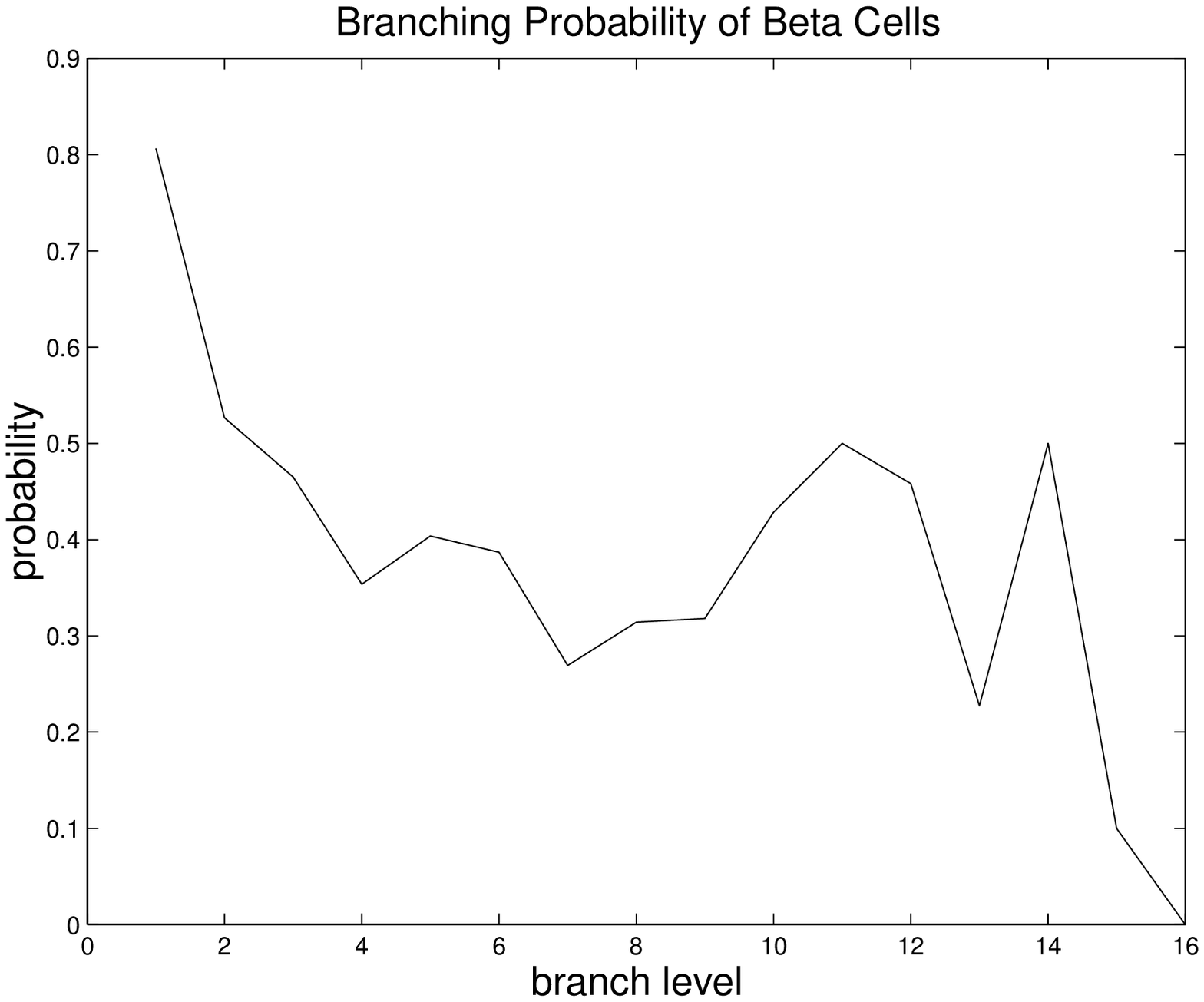}
   \caption{Probability function illustrating the branching probability of alpha (a) and beta (b) cells. ~\label{fig:distr_levels}} \end{center}
\end{figure}

\begin{figure}
 \begin{center} 
   \includegraphics[scale=.4]{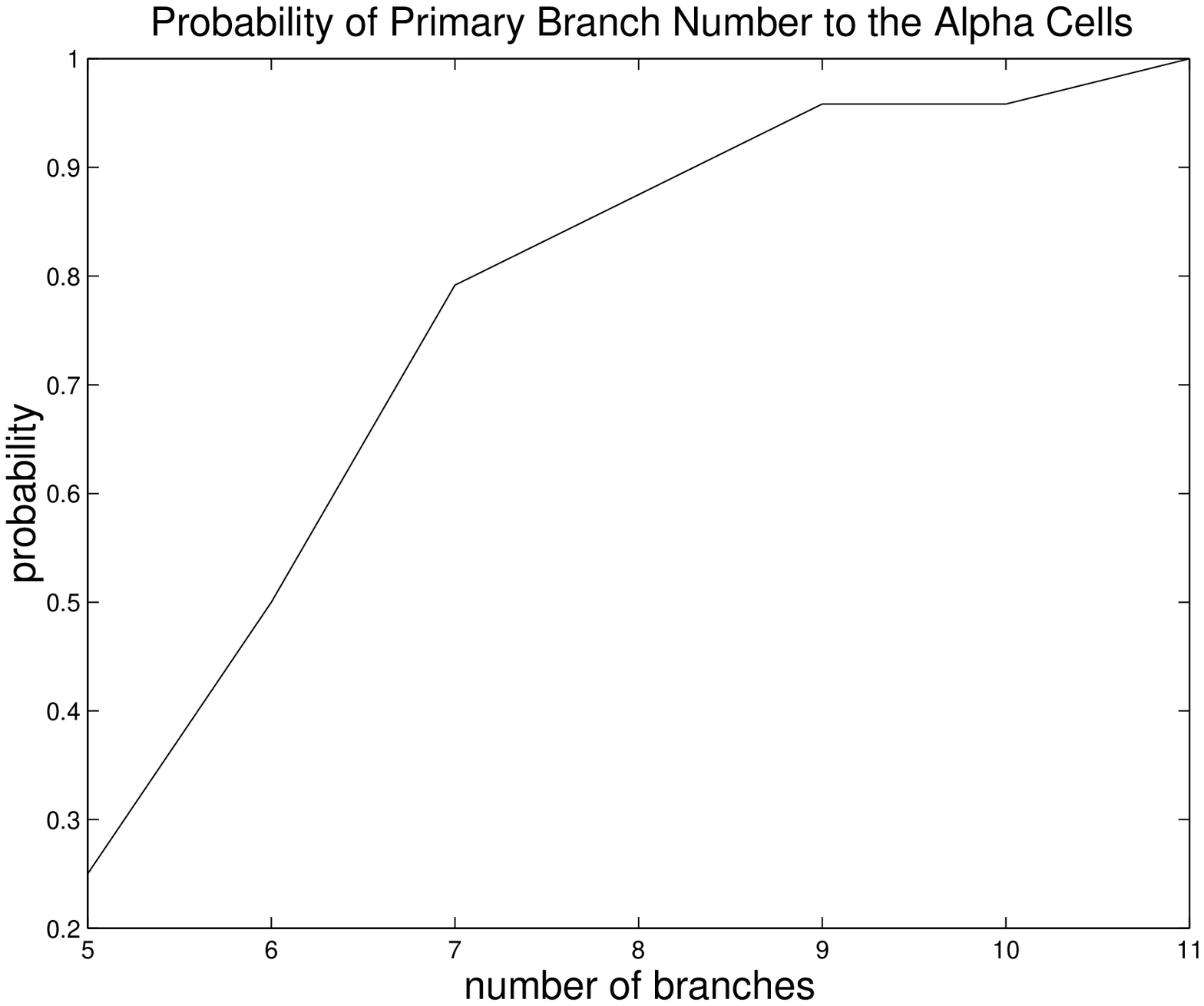}
   \includegraphics[scale=.4]{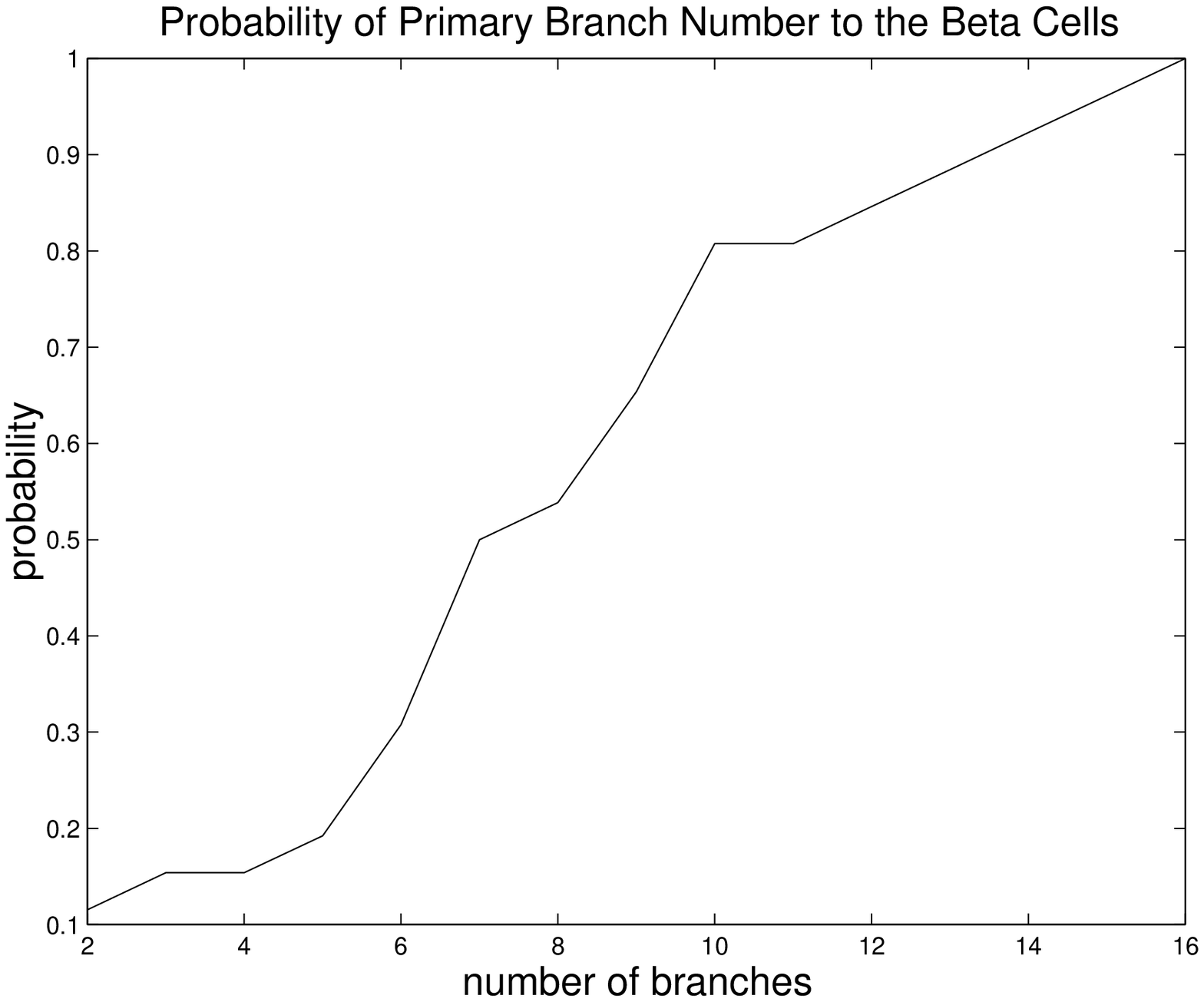}
   \caption{Probability function of the number of primary dendritic segments to the alpha 
(a) and beta (b) cells.~\label{fig:initial}} \end{center}
\end{figure}

The dendrites are henceforth understood as trees, so that the
respective hierarchical level can be precisely defined while
considering the soma as reference.  Therefore, the dendritic segments
directly connected to the soma, as well as branching points initiating
at such segments, are identified as being at hierarchical level 1, and
so on.  Our simulations are restricted to a maximum of 10 hierarchical
levels, as there are very few branchings occurring at higher levels in
the real cells.  The probability of branch points for each considered
type of cell are shown in Figure~\ref{fig:distr_levels}(a) and (b),
respectively for alpha and beta cells.  The probability of the number
of dendritic segments directly attached to the soma (i.e. hierarchical
level 1) is also necessary for the statistical model of the growing
cells.  Figure~\ref{fig:initial} show the cumulative densities for the
alpha (a) and beta (b) neuronal types.

In addition to being essential for neuronal shape modeling, the above
branching and initial densities provide interesting information by
themselves.  For instance, it is clear from the two densities in
Figure~\ref{fig:distr_levels} that the alpha cells are characterized
by higher branching rates at the lower hierarchical levels, as is
clear from the more accentuated decreasal of the respective density
along the hierarchical levels. Figure~\ref{fig:initial} presents the
probability of the number of primary dendritic segments for alpha (a)
and beta (b) cells.  These graphs indicate that the beta cells are
more likely to branch than alpha cells at hierarchical levels between
8 and 12.

\begin{figure}
 \begin{center} 
   \includegraphics[scale=.4]{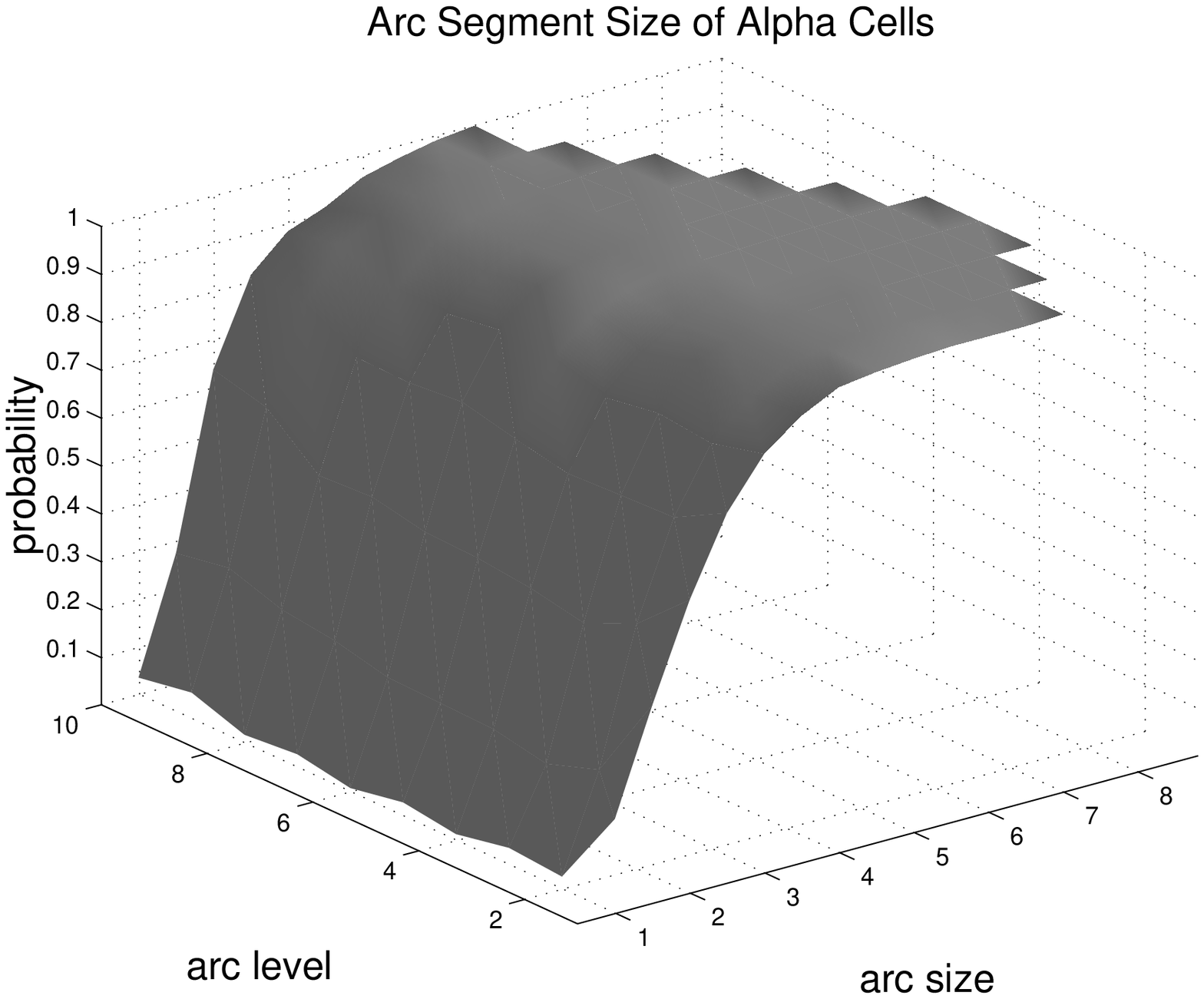}
   \includegraphics[scale=.4]{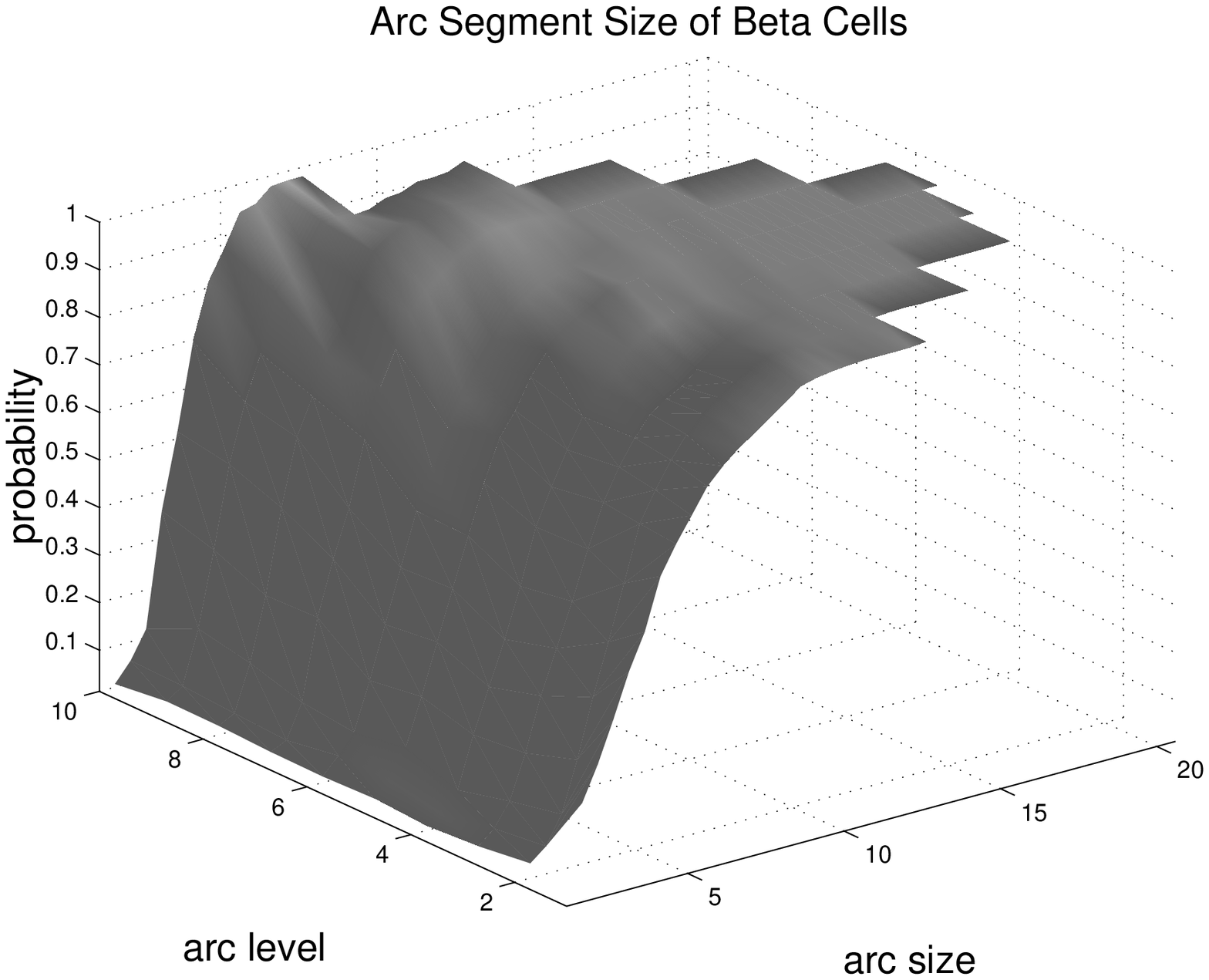} 
   \caption{Distribution function of dendritic segments arc-lengths to the 
   alpha (a) and beta (b) cells.~\label{fig:lengths}} 
  \end{center}
\end{figure}

\begin{figure}
 \begin{center} 
   \includegraphics[scale=.4]{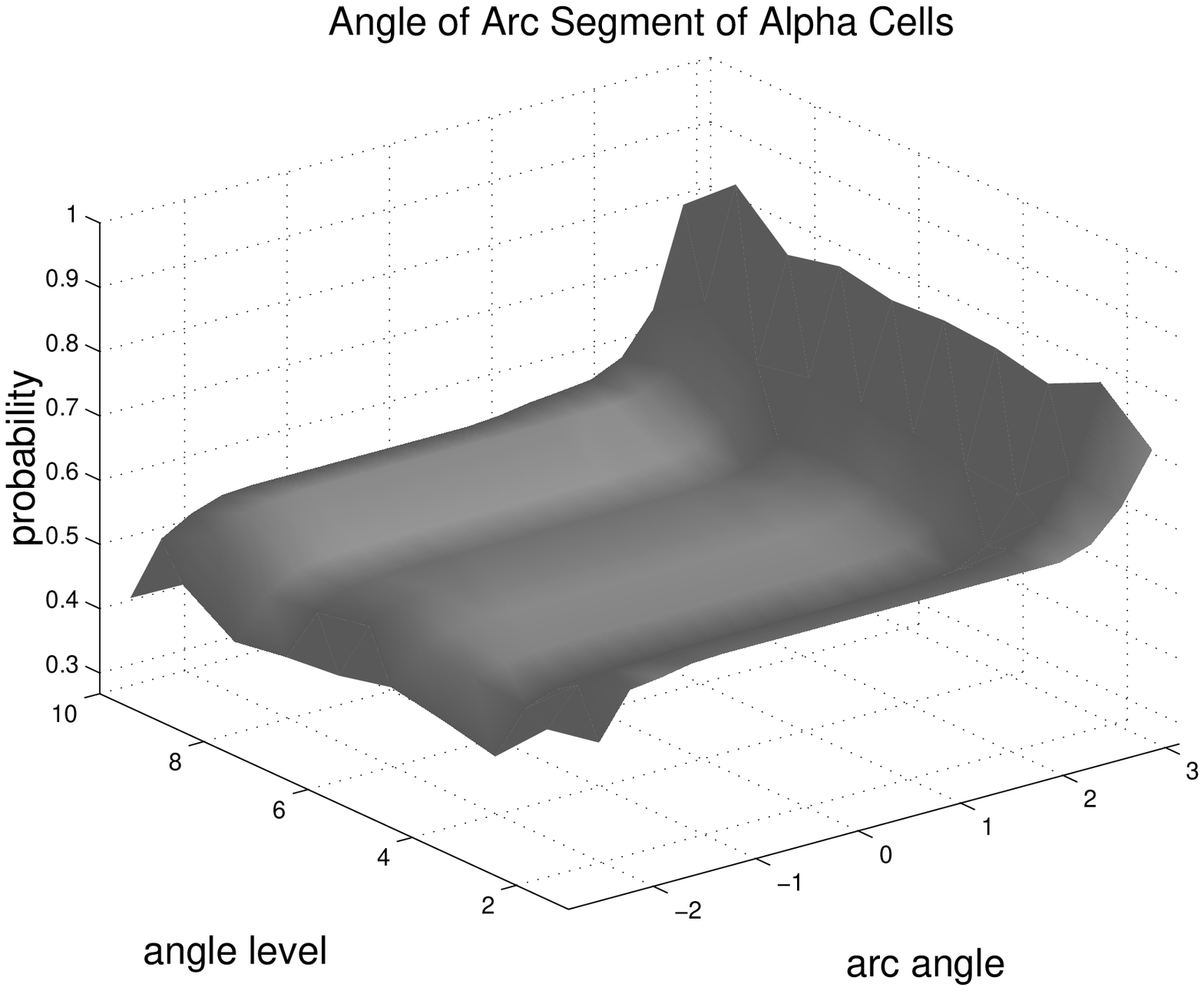}
   \includegraphics[scale=.4]{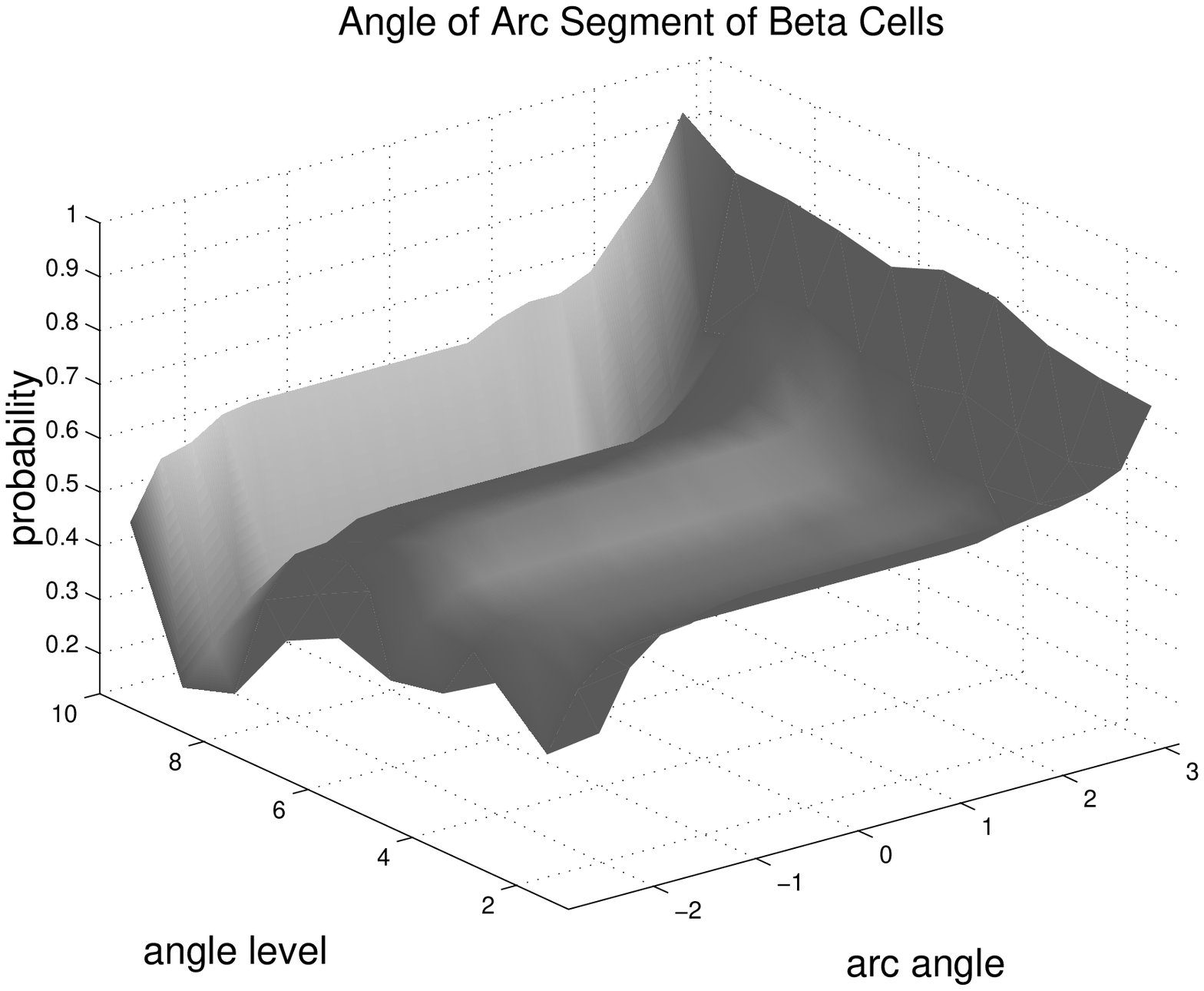}
   \caption{Distribution function of dendritic segments arc-angles to the alpha (a) and beta (b) cells.~\label{fig:arc_angles}}

\end{center}
\end{figure}

\begin{figure}
 \begin{center} 

   \includegraphics[scale=.4]{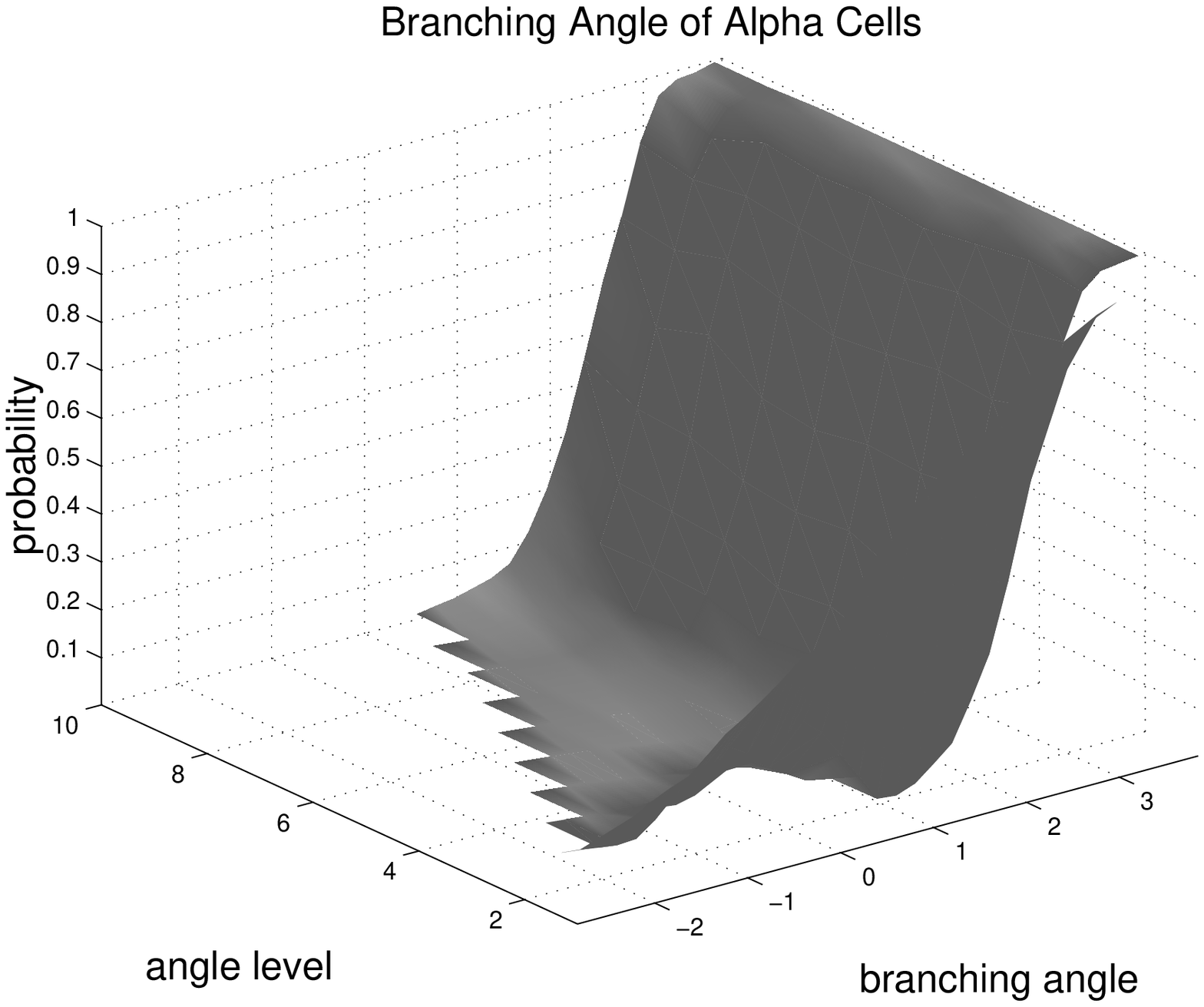}
   \includegraphics[scale=.4]{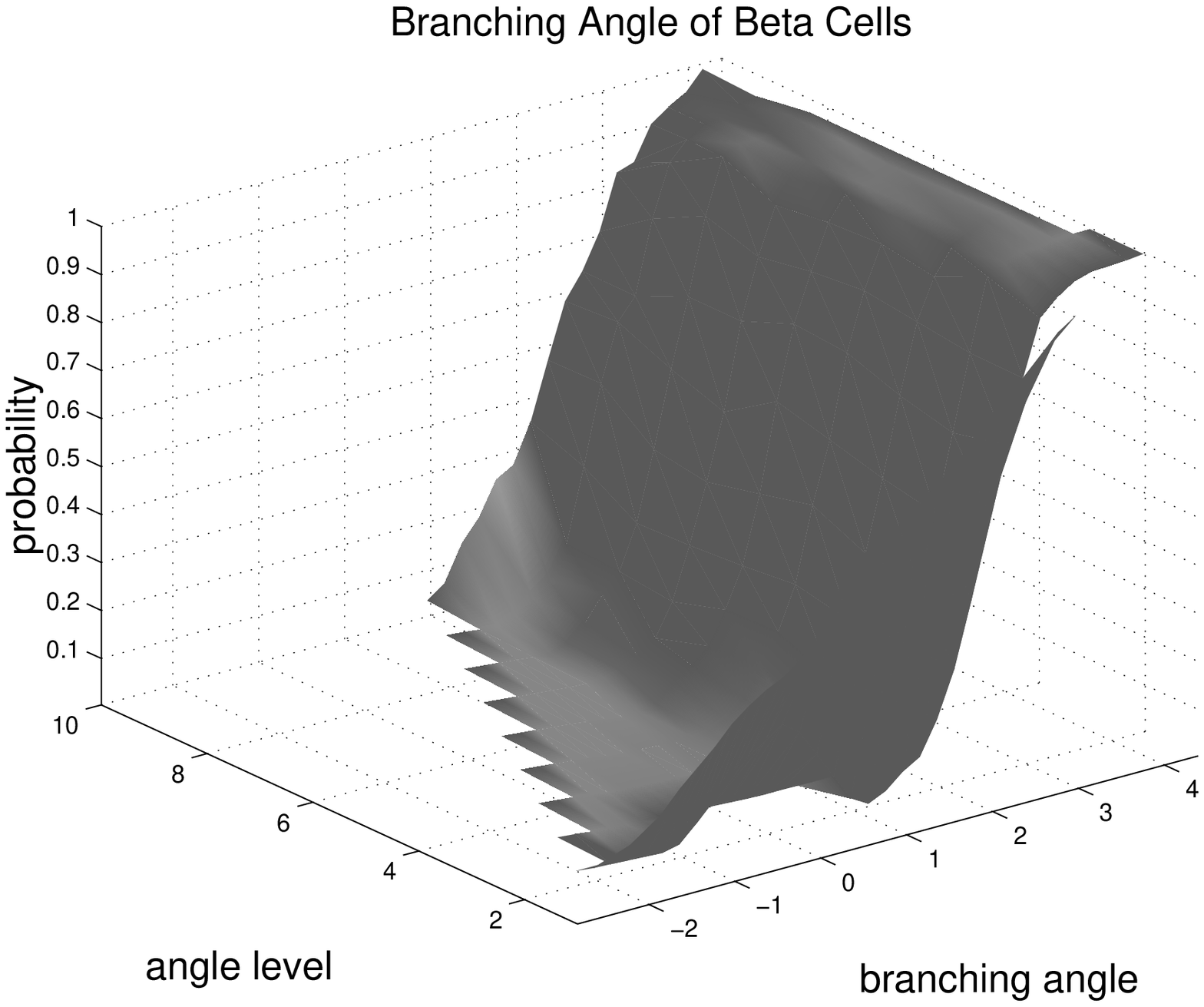}
   \caption{Distribution function of the angles at the branch point to the alpha (a) and beta (b) cells.~\label{fig:angles}}

\end{center}
\end{figure}

\begin{figure}
 \begin{center} 

   \includegraphics[scale=.4]{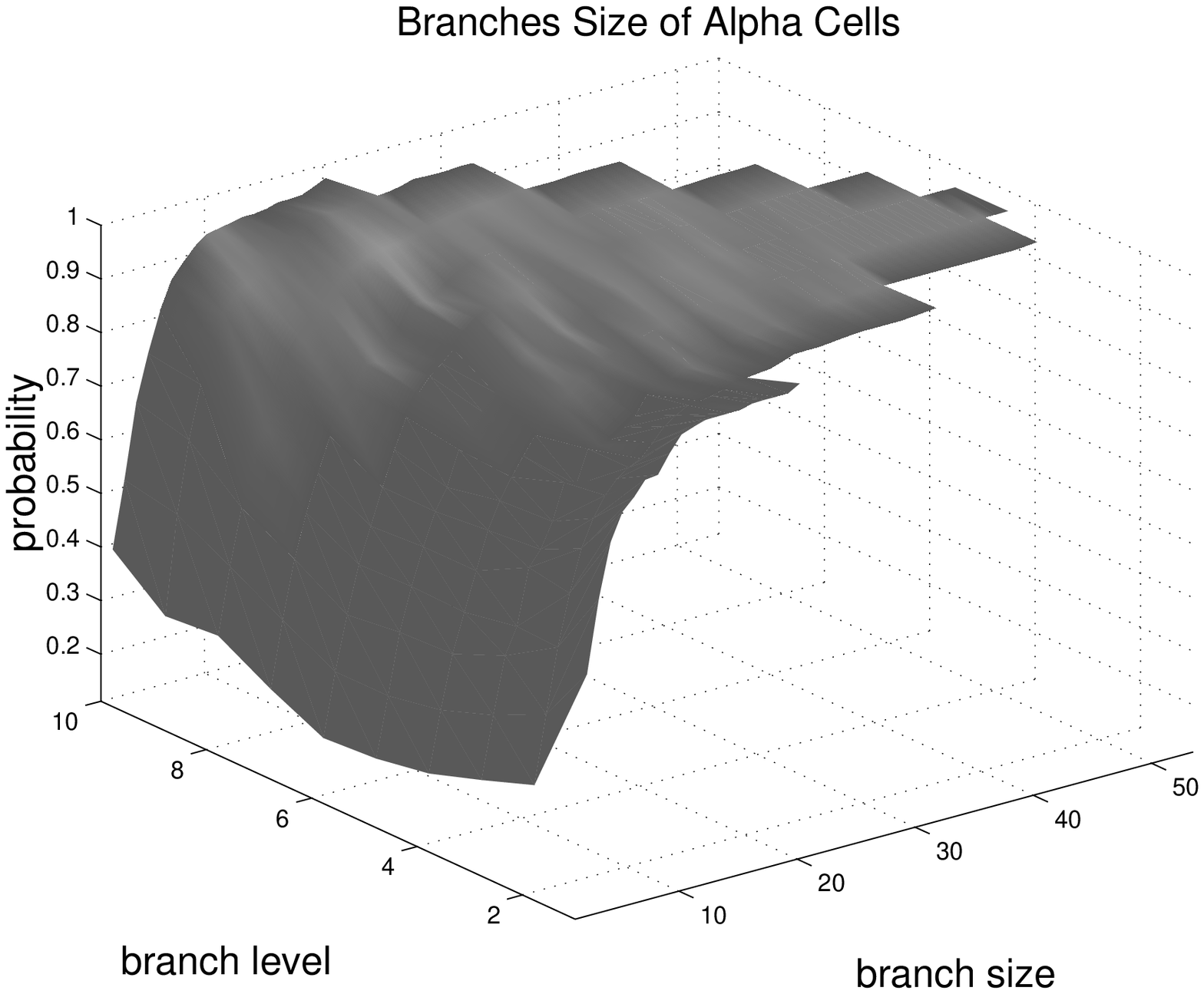}
   \includegraphics[scale=.4]{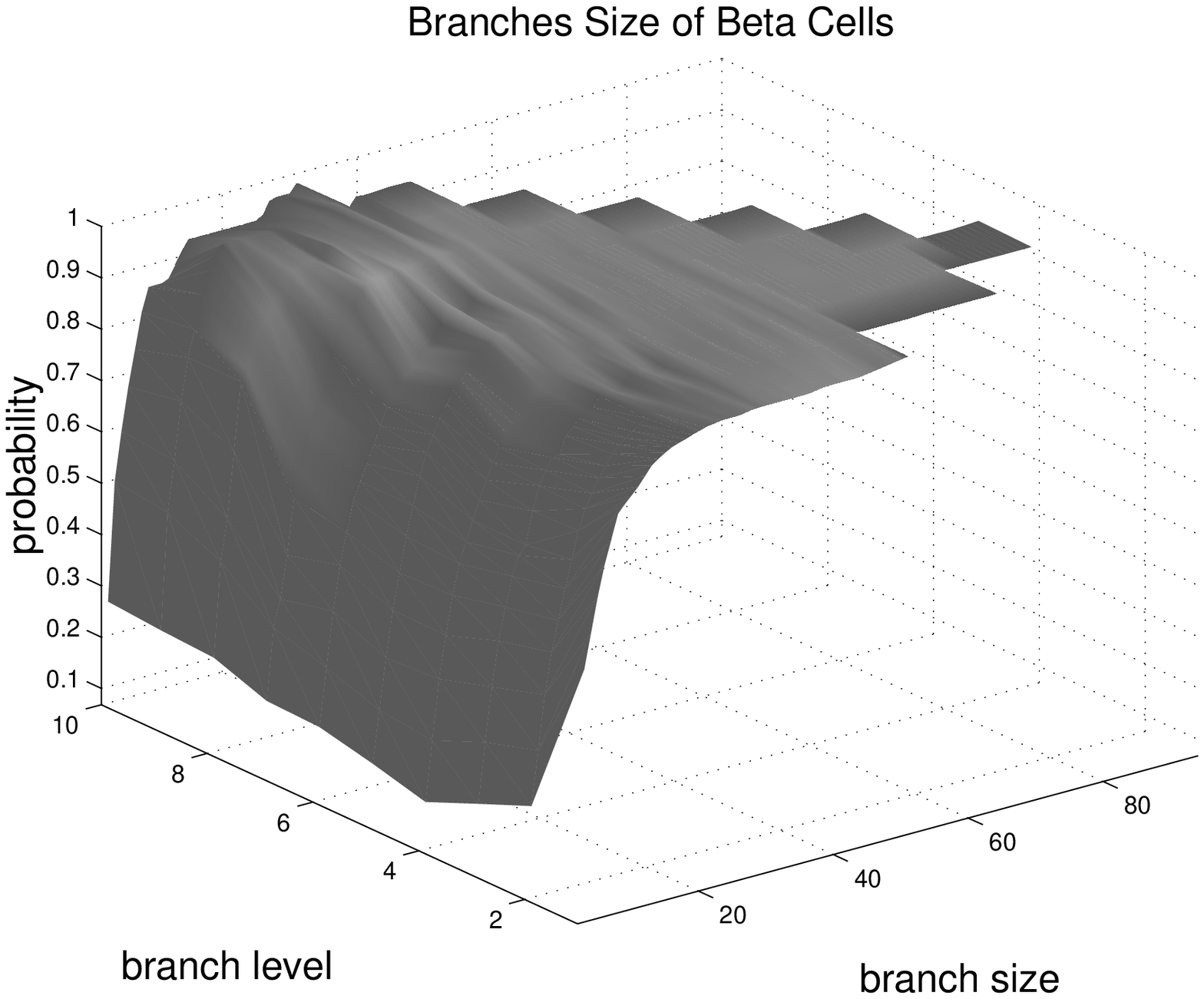}
   \caption{Distribution function of the branch lengths to the alpha 
(a) and beta (b) cells.~\label{fig:branch_lengths}}

\end{center}
\end{figure}

In addition to the above probabilities, it is also necessary to obtain
probabilistic models of the dendritic segment arc-lengths.  Although
alpha cells are typically much larger than beta cells (especially in
the periphery of the retina), we used size-normalized versions of the
considered neurons in order to have neurons with similar sizes.  This
allows our percolation study to be mostly defined by the shape
intrincacy of the cells rather than their sizes~\footnote{Given two
cells with similar morphologies but different sizes, the larger cell
will obviously tend to percolate sooner.}.  Figure~\ref{fig:lengths}
shows the cumulative two-variated density of such lengths in terms of
the hierarchical level for the alpha (a) and beta (b) types of cells,
while Figure~\ref{fig:arc_angles} presents the angles of these
dendritic segment arc-lengths for the alpha (a) and beta(b) cells.
The last features considered in this work refer to the branch lengths
and angles at the branch points, which are shown in
Figures~\ref{fig:angles} and ~\ref{fig:branch_lengths} for the alpha
(a) and beta (b) cells.  By ``branch length'' it is understood the
total arc-length while moving from the branching point to the cell
soma.

Note that both branching angle densities
(i.e. Figures~\ref{fig:arc_angles} and ~\ref{fig:angles}) are similar
for both alpha and beta cells. The length-related densities
(i.e. Figures~\ref{fig:lengths} and ~\ref{fig:branch_lengths}) were
obtained for alpha cells and then normalized with respect to their
respective diameters (i.e. the largest distance between any two points
of each cell) in such a way that they have the same average diameter
as beta cells. Such a normalization was adopted so that the
percolation only reflects the shape (and not size) of the cells.

\section{Neuronal Synthesis}

\begin{figure*}
 \begin{center} 

 \includegraphics[scale=.23]{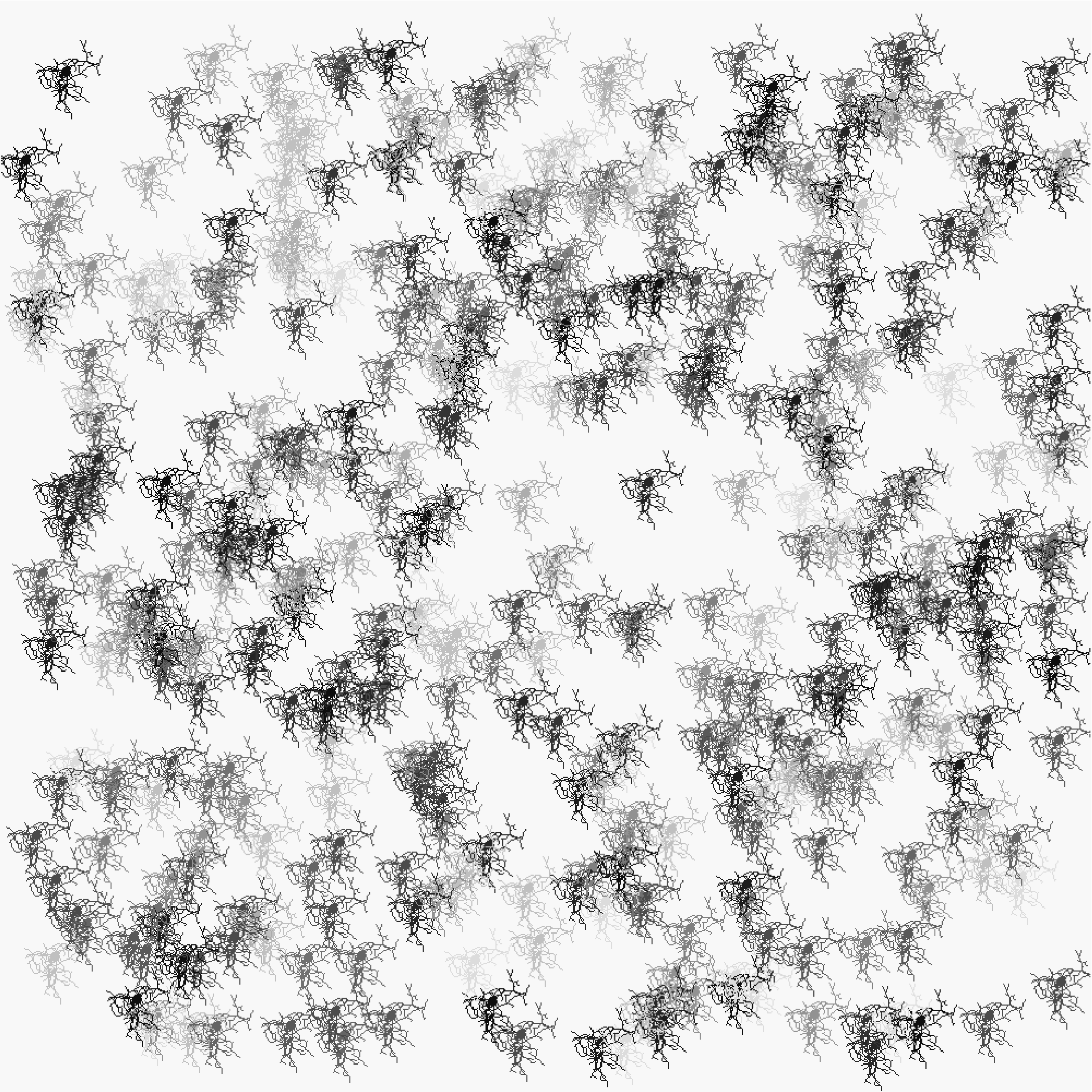} \hspace{1cm}
 \includegraphics[scale=.23]{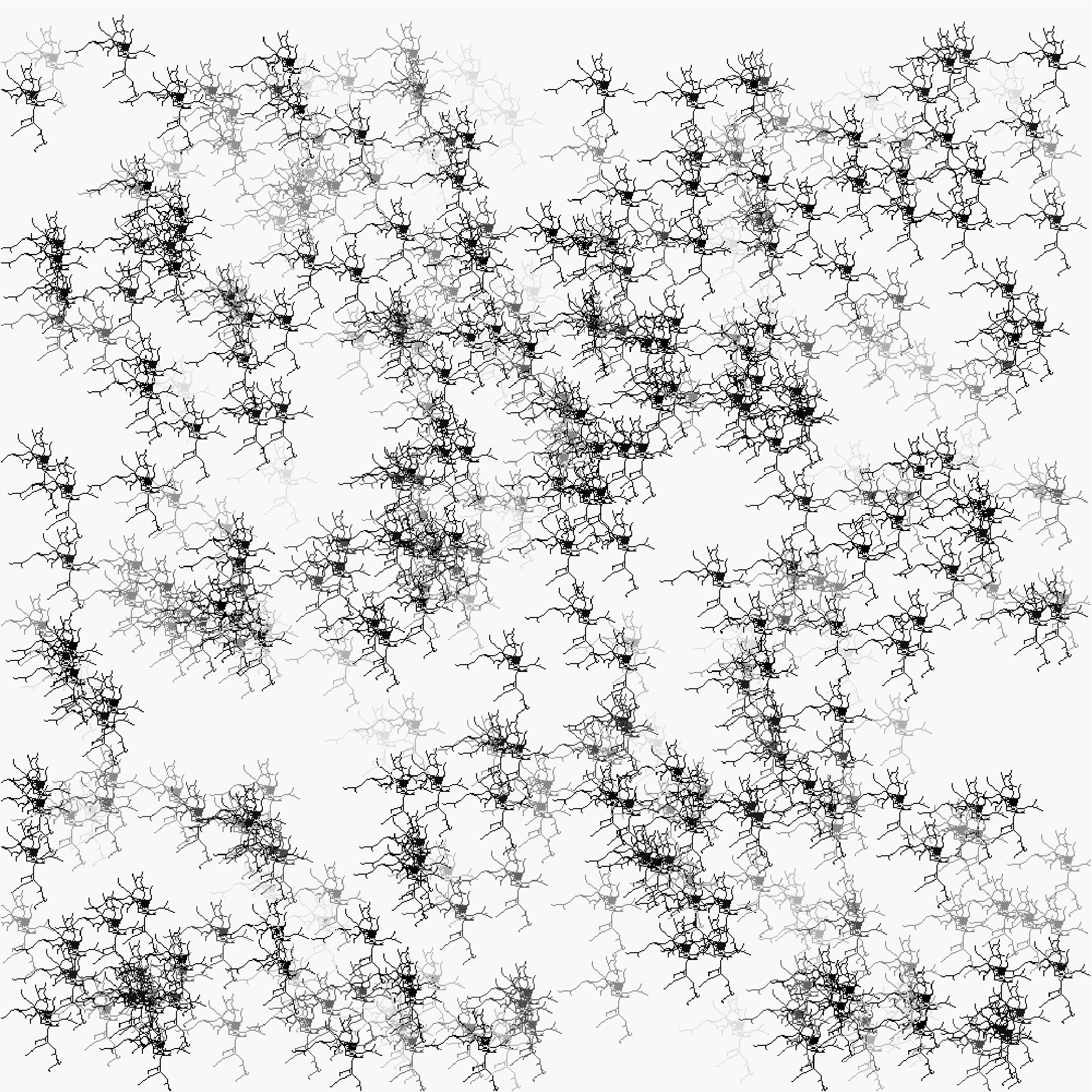} \\ 
 (a) \hspace{9cm}(b) 

  \caption{Examples of neuronal networks obtained by the adopted
  growth methodology considering the alpha (a) and beta (b)
  models. Different gray levels were used for neuronal cell
  representation in order to facilitate the visualization of the
  individual cells.~\label{fig:ex_growth}}

\end{center}
\end{figure*}

In order to generate the neuronal shapes, the probabilistic model of
the neuronal geometry described in the previous section was
statistically sampled by the Monte Carlo approach as explained in the
following.  Initially, the soma of each cell was uniformly (Poisson
distribution) distributed along an $N \times N$ matrix (associated to
a digital image).  The number of branches emerging from the soma was
randomly chosen according to the respective density, being uniformly
distributed along the somata, which are circular.  For each cell, for
each branch, the orientation of the emerging segment was drawn from
the respective distribution. Straight segments are then incorporated,
piece-by-piece, into the growing process.  The length and orientation
of each of these segment pieces was sampled through Monte Carlo from
the respective statistical model, therefore taking into account the
previous angle and length. In order to allow all neuronal cells to
grow in a `simultaneous' fashion, a single segment piece is
incorporated into each growing branch, for each neuronal cell, at a
time (`round-robin' scheme).  Every time a new branch was visited, the
probability for new branch or growth termination was sampled, and the
respective action taken.  In case we have a branch\footnote{A branch
point along the dendritic arborization corresponds to a point where
the growing dendrite bifurctes.}, the orientations of the two
branching new segments were sampled from the respective distributions,
and those branches were subsequently included in the `round-robin'
growth scheme.  The growth of branches continued until one of the
following conditions is reached: (a) it is selected for interruption;
or (b) it reached 10 hierarchical stages.  Figure~\ref{fig:ex_growth}
illustrates morphologically-realistic neuronal networks obtained by
the growing process described above considering alpha (a) and beta (b)
cells.

\section{Percolation Dynamics}

During the simulated neuronal growth, a synaptic connection is
implemented every time a growing dendrite overlaps any portion of the
other current cells.  So, as the cells develop in size and shape, they
tend to form more connections.  A group of connected cells is
henceforth understood as a \emph{cluster}.  A natural representation
of such growing structures can be immediately obtained by using graphs
whose nodes correspond to the neuronal cell soma and the edges
correspond to the synaptic connections.  While several topological and
morphometrical properties of the evolving neuronal networks can be
quantified, in this work attention is concentrated on the size $S(t)$
of the cluster containing the maximum number of nodes --- i.e. the
\emph{dominating cluster} --- found at each time instant $t$ (i.e. the
growing stage).  The sizes $S(t)$ are calculated from the graphs which
are constructed as the networks evolve.  The critical phenomenon of
percolation is identified by looking for an abrupt transition along
$S(t)$, which is related to the formation of the giant cluster
\cite{Book_Stauffer}.  After this point, the growing neuronal
structure is characterized by the presence of such a giant community,
which dominates the subsequent connectivity dynamics.

A single model of cell was considered in each simulation in order to
keep statistical variability low and allow a more precise
identification of the percolation critical point (not a density as in
traditional percolation theory, but a time instant during the neuronal
outgrowth). The chosen neuronal model is ``stamped'' $N$ times on the
considered space (a rectangular window of 1000 by 1000 elements)
according to the uniform probability.  A total of 500 realizations was
performed for each considered configuration, from which the average
and standard deviation shown in the graphs were obtained.  In order to
avoid intense superposition between cells, cells were placed at least
5 pixels apart one another.

\begin{figure*}
 \begin{center} 

   \includegraphics[scale=.37]{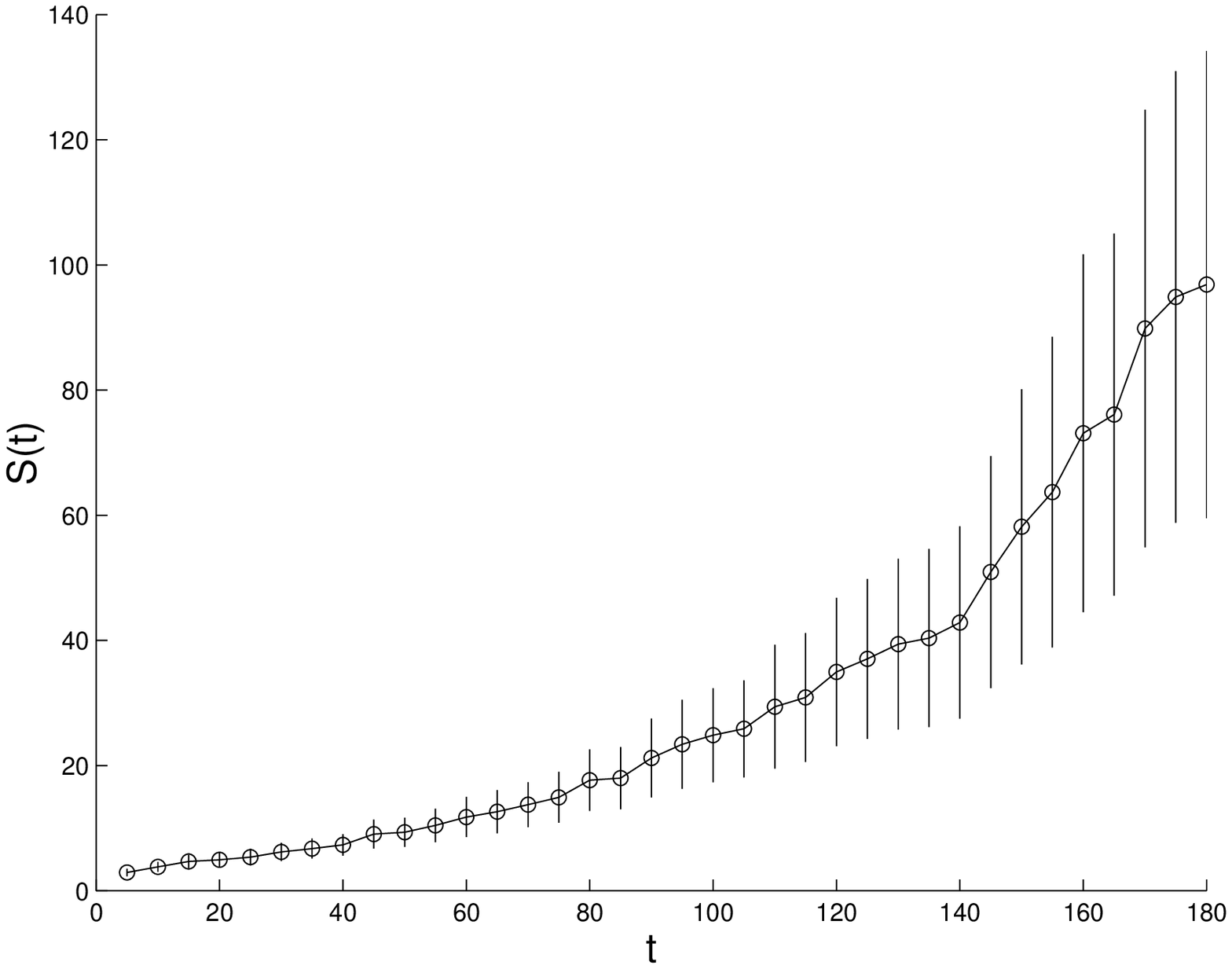}  \hspace{2cm}
   \includegraphics[scale=.37]{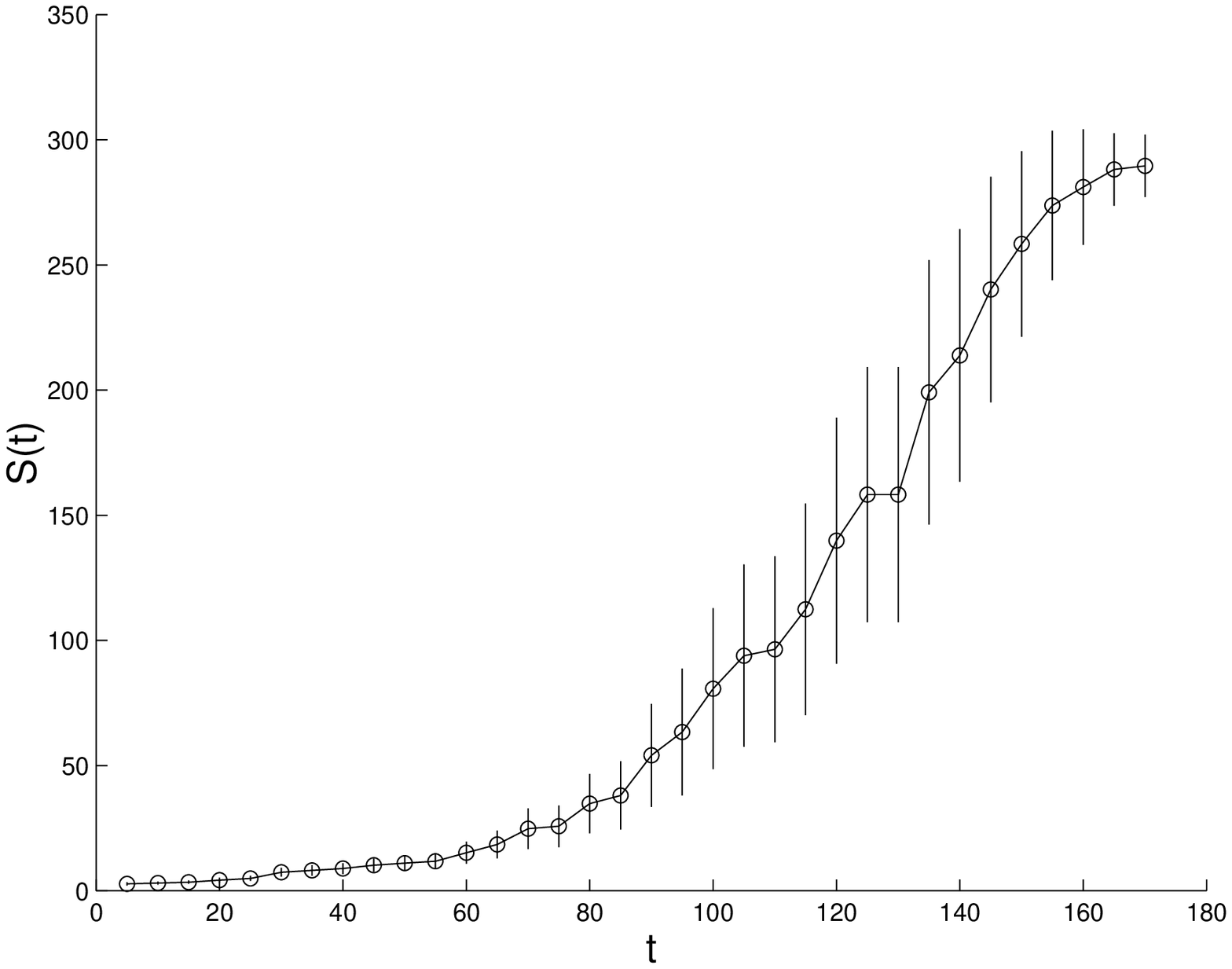} \\
   (a)  \hspace{8cm}  (e) \\

   \includegraphics[scale=.37]{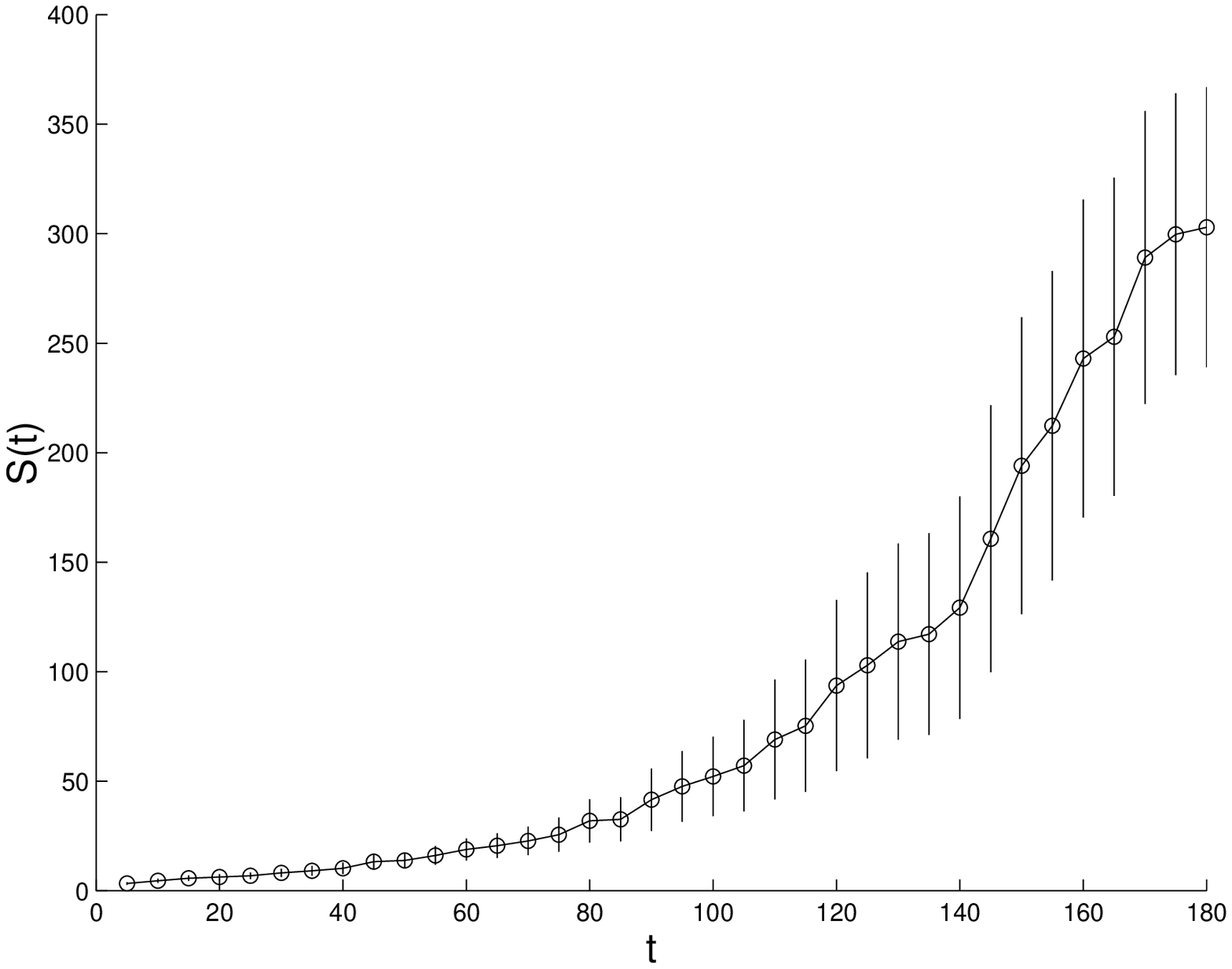}  \hspace{2cm}
   \includegraphics[scale=.37]{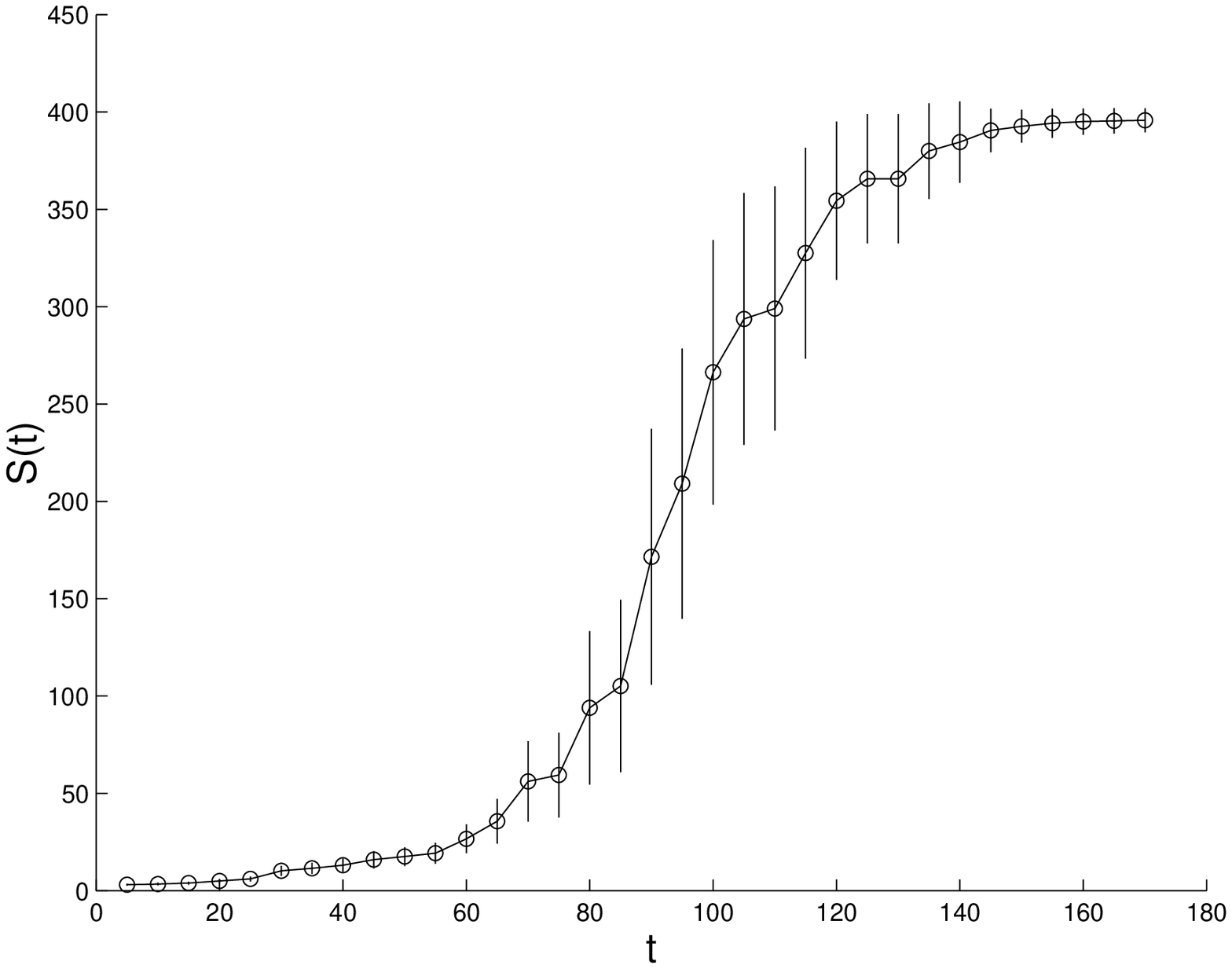} \\
   (b)  \hspace{8cm}  (f) \\

   \includegraphics[scale=.37]{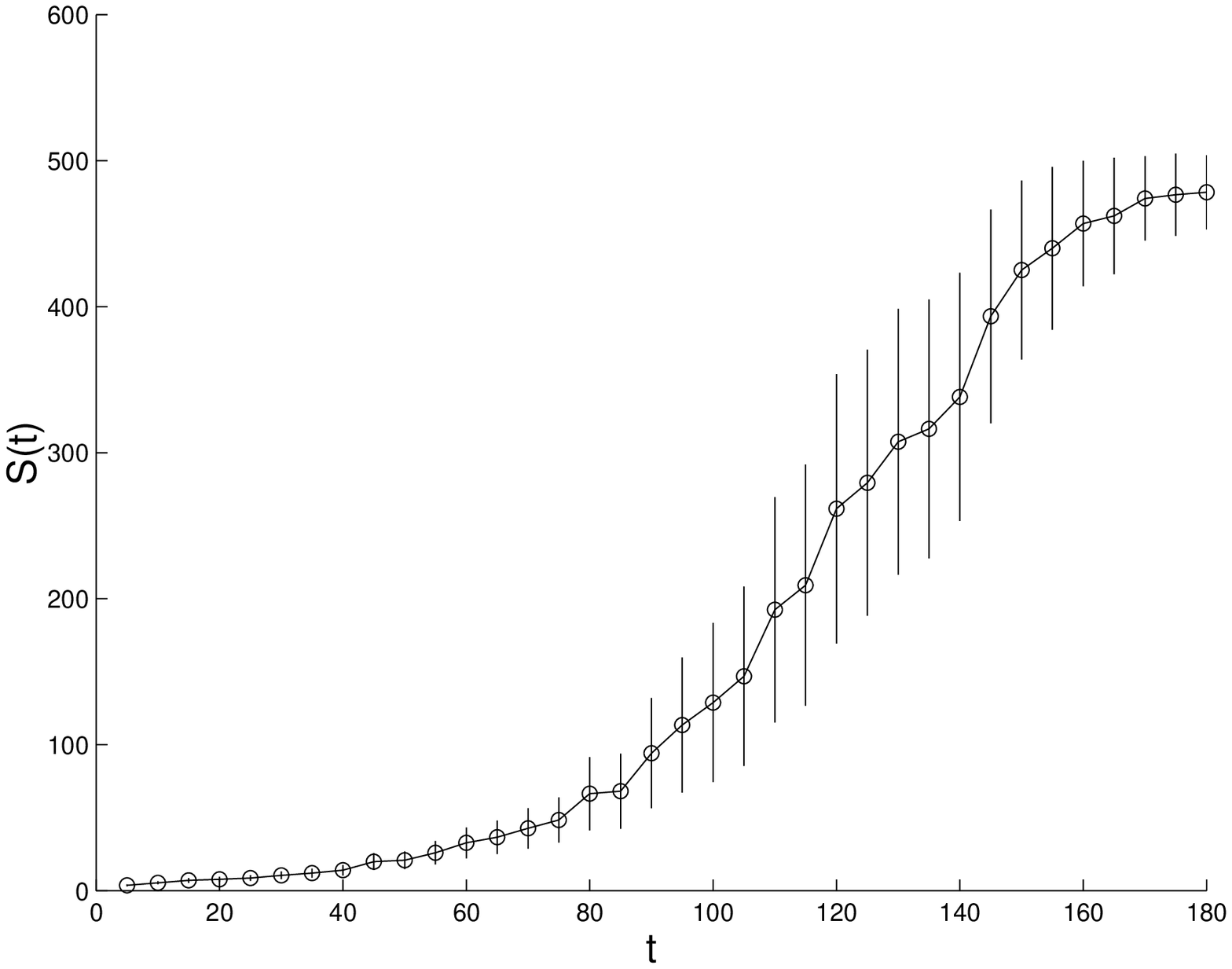}  \hspace{2cm}
   \includegraphics[scale=.37]{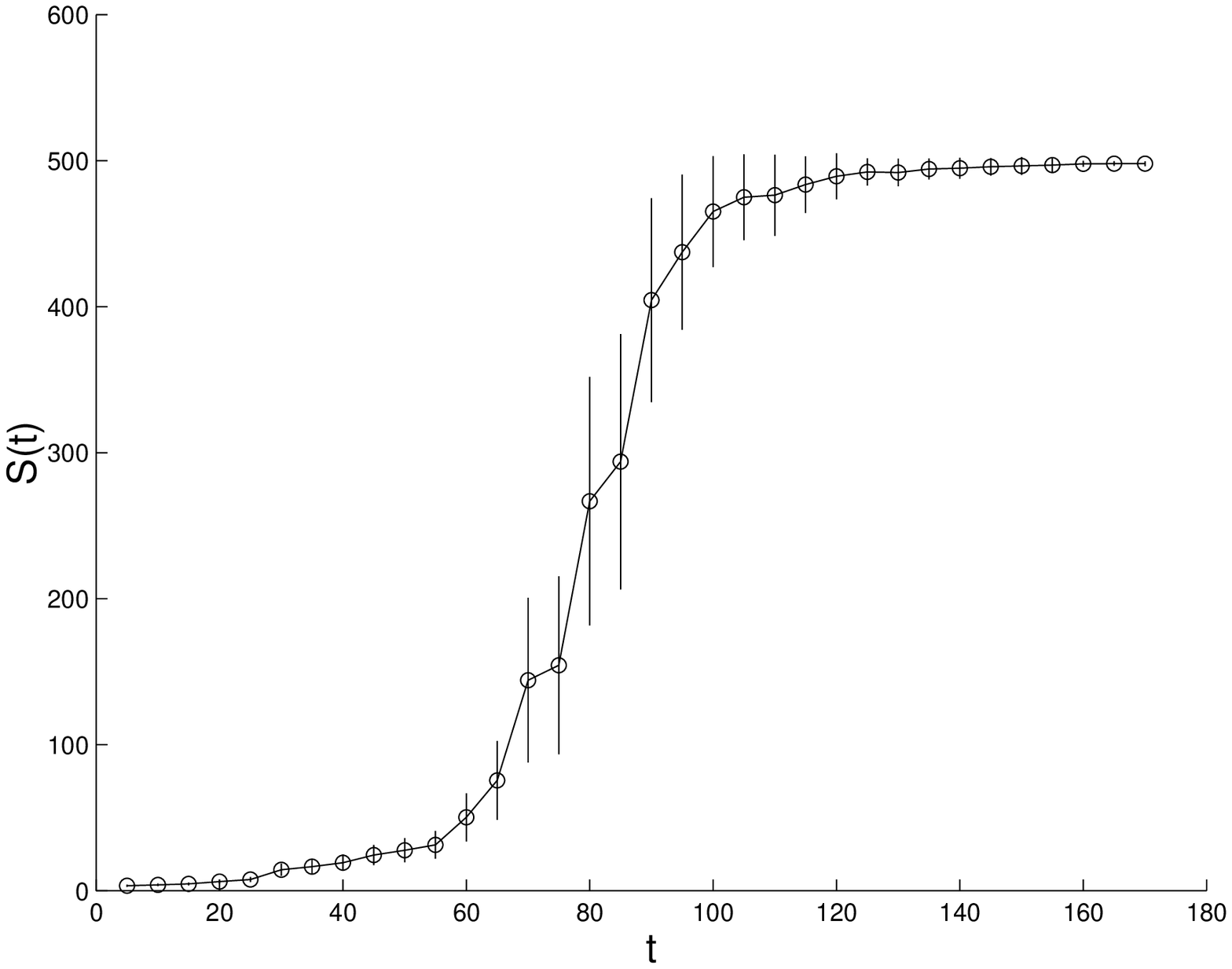} \\
   (c)  \hspace{8cm}  (g) \\

   \includegraphics[scale=.37]{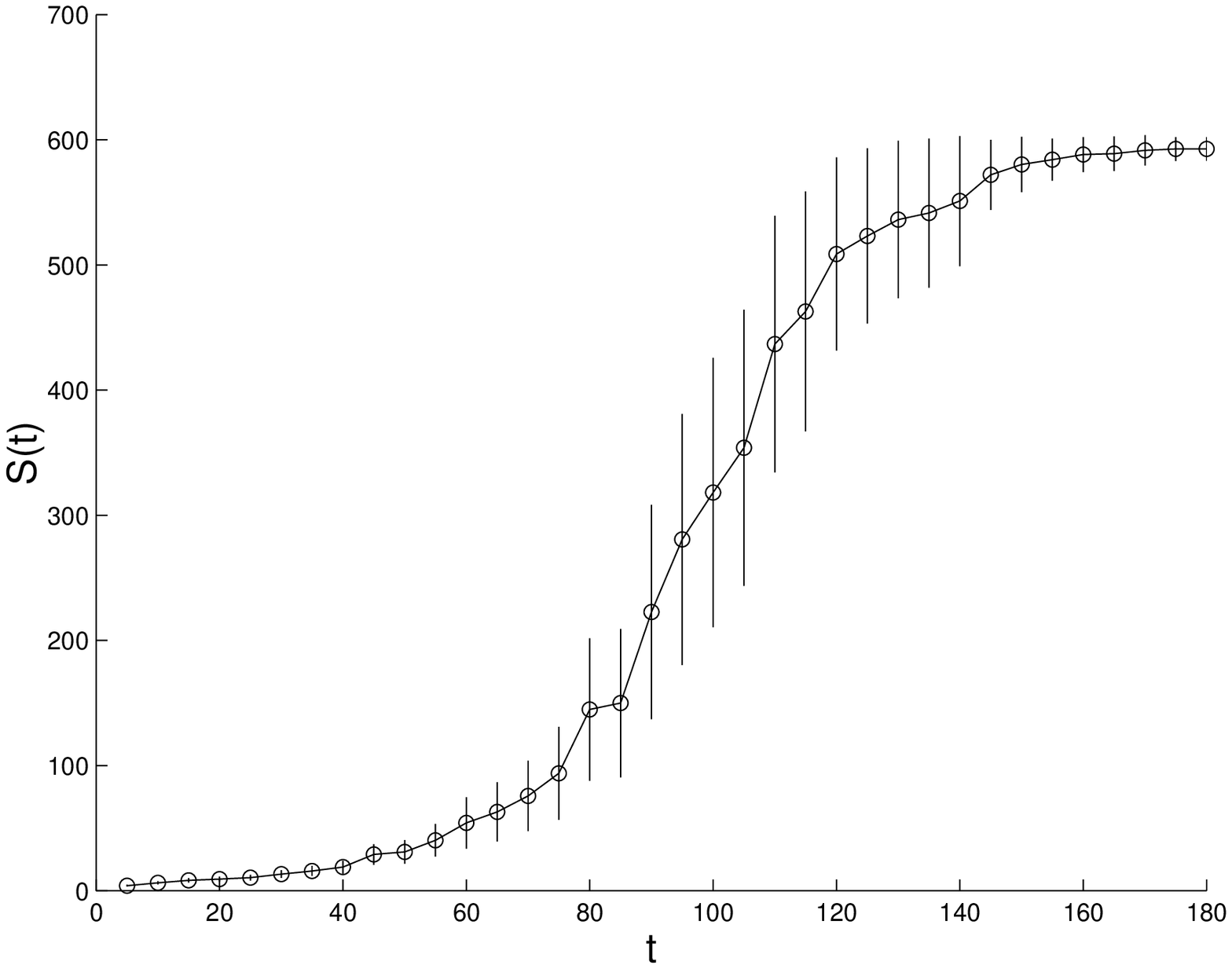}  \hspace{2cm}
   \includegraphics[scale=.37]{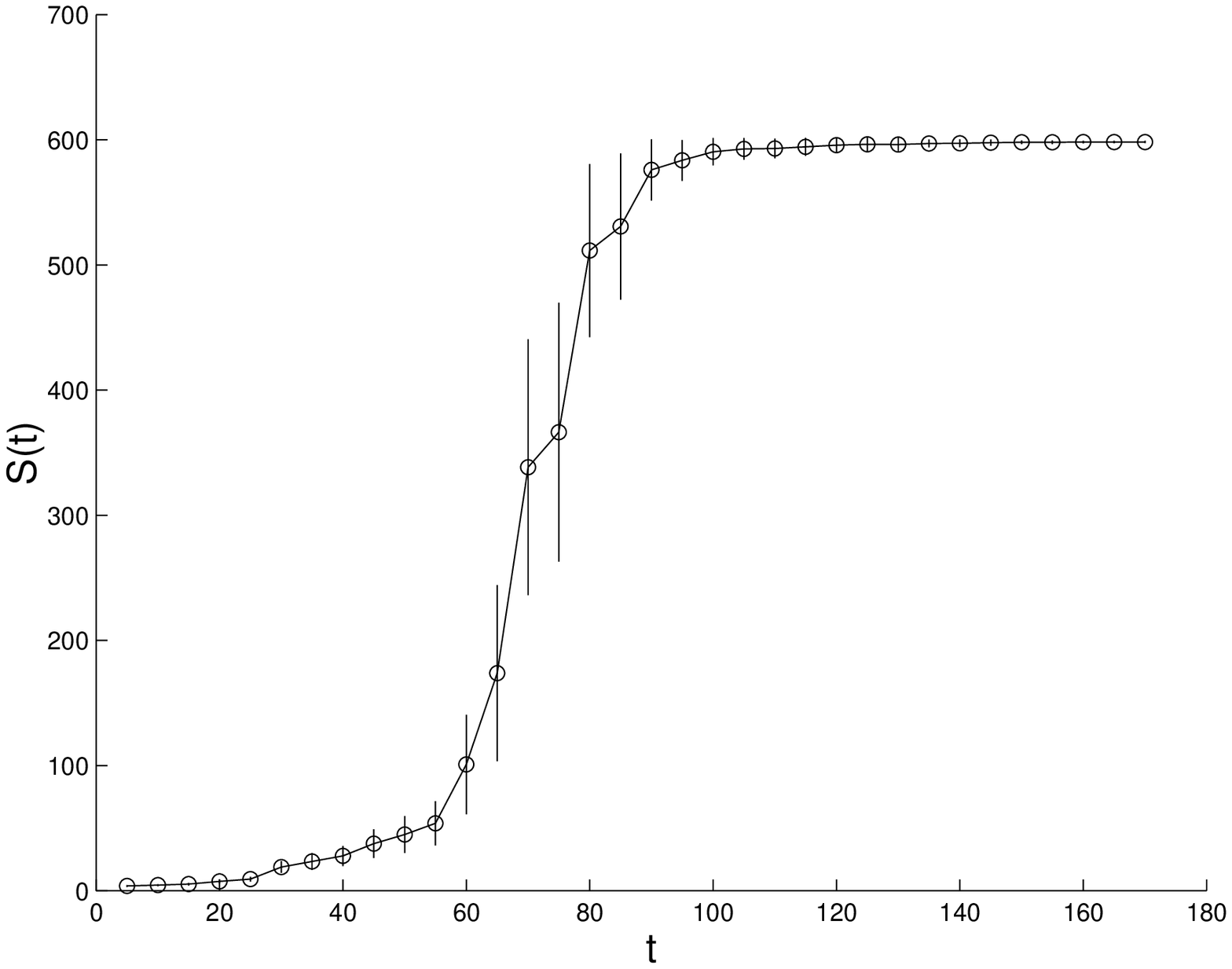} \\
   (d)  \hspace{8cm}  (h)

   \caption{Mean and standard deviation of the size of the largest
cluster in terms of increasing densities (300, 400, 500 and 600 cells)
of alpha (a-d) and beta (e-h)
cells.~\label{fig:connected_group_alpha}}

\end{center}
\end{figure*}

Figure~\ref{fig:connected_group_alpha}(a-d) presents the evolution of
the maximum cluster size considering growing densities of alpha cells,
while Figure~\ref{fig:connected_group_alpha}(e-h) presents analogous
graphs considering beta cells. As expected, the critical transition
tends to increase with the density of neurons, with markedly sharper
transition being verified for the beta neuronal cells.  More
interestingly, it is also clear from the obtained results that the
alpha cells implied longer times (e.g. higher cell densities) before
percolation.  Generally, percolation was often observed after 400
growth steps for beta cells, but only after 600 steps for alpha cells.
This is exactly the opposite result than it would be obtained in case
the cells were not size-normalized.  Indeed, the fact that
size-normalized beta cells tended to percolate sooner than alpha cells
provides a clear indication that such cells tend to have more
intrincated morphology.  As indicated in Figure~\ref{fig:arc_angles},
which describes the cumulative two-variated distribution of dendritic
segment angles, beta cells are characterized by higher dispersion of
angles, implying the overall dendrites to become more disordered and
spatially complex, which is in full agreement with the obtained
percolation dynamics, i.e. more complex neuronal cells tend to
percolate sooner than less complex cells with similar sizes.

\section{Concluding Remarks}

This article has reported on several new perspectives related to
neuromorphic models and percolation induced by dendritic growth.
First, we have shown how morphologically realistic neuronal networks
can be simulated by using Monte-Carlo sampling of statistical models
derived from a series of geometrical measurements of real neuronal
cells.  Second, we have investigated a new perspective to percolation
studies in which, instead of incorporating new connections of fixed
size between the involved elements, the percolation dynamics is
defined by the progressive growth of dendrites/axons, following
biologically-realistic rules derived from experimental data.  The
obtained results indicate that the percolation in such evolving
systems is also characterized by abrupt transitions of the dominating
cluster size along the progression of the growth and connections.  We
have shown that distinct critical points are usually identified for
growing dynamics of systems underlain by distinct neuronal
morphologies, with beta cells reaching percolation sooner than alpha
cells, a result that is related to the fact that the dendritic
processes of beta cells are more intrincated and spatially complex.

Such results establish and interesting connection between the
statistical geometrical features of the considered cells and their
potential for forming clusters among the neuronal milieu.  Such
perspectives and results are particularly interesting because the
functional properties of neuronal networks are closely related to
their connectivity (e.g. \cite{Stauffer_Costa:2003, LU101,
Costa_Velte:1999, Costa_BM:2003}).  The perspectives for further
investigations are many.  For instance, it would be particularly
interesting to check how the consideration of more than one distinct
statistical model of neuronal geometry will affect the measured
critical point.  Another interesting possibility is to investigate, in
the spirit of \cite{LU101}, to what an extent the critical point
statistics can be used as a resource for classification of the
morphological types of involved neuronal cells.  A third promising
future development is to quantify, through simulations, how the
geometrical properties of the neuronal cells
(e.g. \cite{Costa_Velte:1999}), by controlling the sizes of the
neuronal clusters, ultimately define the functional properties of the
obtained structures \cite{Costa_BM:2003}.  Still, it would be
interesting to correlate several measurements from complex network
research, especially those related to the hierarchical structure of
the networks \cite{Costa_hier:2004}, with the critical percolation
time.

\begin{acknowledgments}
Luciano da F. Costa is grateful to FAPESP (processes 99/12765-2) and
CNPq (process 308231/03-1) for financial support.  The authors thank
Luis Diambra, Marconi S. Barbosa and Gonzalo Travieso for reviewing
and commenting on this work.
\end{acknowledgments}

\bibliography{pre_reg}

\begin{thebibliography}{20}
\expandafter\ifx\csname natexlab\endcsname\relax\def\natexlab#1{#1}\fi
\expandafter\ifx\csname bibnamefont\endcsname\relax
  \def\bibnamefont#1{#1}\fi
\expandafter\ifx\csname bibfnamefont\endcsname\relax
  \def\bibfnamefont#1{#1}\fi
\expandafter\ifx\csname citenamefont\endcsname\relax
  \def\citenamefont#1{#1}\fi
\expandafter\ifx\csname url\endcsname\relax
  \def\url#1{\texttt{#1}}\fi
\expandafter\ifx\csname urlprefix\endcsname\relax\def\urlprefix{URL }\fi
\providecommand{\bibinfo}[2]{#2}
\providecommand{\eprint}[2][]{\url{#2}}

\bibitem[{\citenamefont{Kandel et~al.}(1995)\citenamefont{Kandel, Schwartz, and
  Jessel}}]{Kandel:1995}
\bibinfo{author}{\bibfnamefont{E.~R.} \bibnamefont{Kandel}},
  \bibinfo{author}{\bibfnamefont{J.~H.} \bibnamefont{Schwartz}},
  \bibnamefont{and} \bibinfo{author}{\bibfnamefont{T.~M.}
  \bibnamefont{Jessel}}, \emph{\bibinfo{title}{Essentials of neural science and
  behavior}} (\bibinfo{publisher}{Appleton and Lange},
  \bibinfo{address}{Englewood Cliffs}, \bibinfo{year}{1995}).

\bibitem[{\citenamefont{Albert and Barab\'asi}(2002)}]{Albert_Barab:2002}
\bibinfo{author}{\bibfnamefont{R.}~\bibnamefont{Albert}} \bibnamefont{and}
  \bibinfo{author}{\bibfnamefont{A.~L.} \bibnamefont{Barab\'asi}},
  \bibinfo{journal}{Rev. Mod. Phys.} \textbf{\bibinfo{volume}{74}},
  \bibinfo{pages}{47} (\bibinfo{year}{2002}).

\bibitem[{\citenamefont{Newman}(2003)}]{Newman:2003}
\bibinfo{author}{\bibfnamefont{M.~E.~J.} \bibnamefont{Newman}},
  \bibinfo{journal}{SIAM Review} \textbf{\bibinfo{volume}{45}},
  \bibinfo{pages}{167} (\bibinfo{year}{2003}),
  \bibinfo{note}{cond-mat/0303516}.

\bibitem[{\citenamefont{Dorogovtsev and Mendes}(2002)}]{Dorog_Mendes:2002}
\bibinfo{author}{\bibfnamefont{S.~N.} \bibnamefont{Dorogovtsev}}
  \bibnamefont{and} \bibinfo{author}{\bibfnamefont{J.~F.~F.}
  \bibnamefont{Mendes}}, \bibinfo{journal}{Advances in Physics}
  \textbf{\bibinfo{volume}{51}}, \bibinfo{pages}{1079} (\bibinfo{year}{2002}),
  \bibinfo{note}{cond-mat/0106144}.

\bibitem[{\citenamefont{Stauffer and Aharony}(1991)}]{Book_Stauffer}
\bibinfo{author}{\bibfnamefont{D.}~\bibnamefont{Stauffer}} \bibnamefont{and}
  \bibinfo{author}{\bibfnamefont{A.}~\bibnamefont{Aharony}},
  \emph{\bibinfo{title}{An introduction to percolation theory}}
  (\bibinfo{publisher}{Taylor and Francis}, \bibinfo{year}{1991}),
  \bibinfo{note}{second edition}.

\bibitem[{\citenamefont{Karbowski}(2003)}]{Karbowski:2001}
\bibinfo{author}{\bibfnamefont{J.}~\bibnamefont{Karbowski}},
  \bibinfo{journal}{Phys. Rev. Lett.} \textbf{\bibinfo{volume}{86}},
  \bibinfo{pages}{3674} (\bibinfo{year}{2003}).

\bibitem[{\citenamefont{Shefi et~al.}(2002)\citenamefont{Shefi, Golding, Segev,
  Ben-Jacob, and Ayali}}]{Shefi:2002}
\bibinfo{author}{\bibfnamefont{O.}~\bibnamefont{Shefi}},
  \bibinfo{author}{\bibfnamefont{I.}~\bibnamefont{Golding}},
  \bibinfo{author}{\bibfnamefont{R.}~\bibnamefont{Segev}},
  \bibinfo{author}{\bibfnamefont{E.}~\bibnamefont{Ben-Jacob}},
  \bibnamefont{and} \bibinfo{author}{\bibfnamefont{A.}~\bibnamefont{Ayali}},
  \bibinfo{journal}{Phys. Rev. E} \textbf{\bibinfo{volume}{66}},
  \bibinfo{pages}{021905} (\bibinfo{year}{2002}).

\bibitem[{\citenamefont{da~F.~Costa and Monteiro}(2003)}]{Percolation:2003}
\bibinfo{author}{\bibfnamefont{L.}~\bibnamefont{da~F.~Costa}} \bibnamefont{and}
  \bibinfo{author}{\bibfnamefont{E.~T.~M.} \bibnamefont{Monteiro}},
  \bibinfo{journal}{Neuroinformatics} \textbf{\bibinfo{volume}{1}},
  \bibinfo{pages}{65} (\bibinfo{year}{2003}).

\bibitem[{\citenamefont{Segev et~al.}(2003)\citenamefont{Segev, Benveniste,
  Shapira, and Ben-Jacob}}]{Segev:2003}
\bibinfo{author}{\bibfnamefont{R.}~\bibnamefont{Segev}},
  \bibinfo{author}{\bibfnamefont{M.}~\bibnamefont{Benveniste}},
  \bibinfo{author}{\bibfnamefont{Y.}~\bibnamefont{Shapira}}, \bibnamefont{and}
  \bibinfo{author}{\bibfnamefont{E.}~\bibnamefont{Ben-Jacob}},
  \bibinfo{journal}{Phys. Rev. Lett.} \textbf{\bibinfo{volume}{90}},
  \bibinfo{pages}{168101} (\bibinfo{year}{2003}).

\bibitem[{\citenamefont{Stauffer et~al.}(2003)\citenamefont{Stauffer, Aharony,
  da~F.~Costa, and Adler}}]{Stauffer_Costa:2003}
\bibinfo{author}{\bibfnamefont{D.}~\bibnamefont{Stauffer}},
  \bibinfo{author}{\bibfnamefont{A.}~\bibnamefont{Aharony}},
  \bibinfo{author}{\bibfnamefont{L.}~\bibnamefont{da~F.~Costa}},
  \bibnamefont{and} \bibinfo{author}{\bibfnamefont{J.}~\bibnamefont{Adler}},
  \bibinfo{journal}{Eur. J. Phys. B} \textbf{\bibinfo{volume}{32}},
  \bibinfo{pages}{395} (\bibinfo{year}{2003}).

\bibitem[{\citenamefont{da~F.~Costa and Stauffer}(2003)}]{Costa_Stauffer:2003}
\bibinfo{author}{\bibfnamefont{L.}~\bibnamefont{da~F.~Costa}} \bibnamefont{and}
  \bibinfo{author}{\bibfnamefont{D.}~\bibnamefont{Stauffer}},
  \bibinfo{journal}{Physica A} \textbf{\bibinfo{volume}{330}}
  (\bibinfo{year}{2003}).

\bibitem[{\citenamefont{da~F.~Costa et~al.}(2003)\citenamefont{da~F.~Costa,
  Barbosa, Coupez, and Stauffer}}]{Costa_BM:2003}
\bibinfo{author}{\bibfnamefont{L.}~\bibnamefont{da~F.~Costa}},
  \bibinfo{author}{\bibfnamefont{M.~S.} \bibnamefont{Barbosa}},
  \bibinfo{author}{\bibfnamefont{V.}~\bibnamefont{Coupez}}, \bibnamefont{and}
  \bibinfo{author}{\bibfnamefont{D.}~\bibnamefont{Stauffer}},
  \bibinfo{journal}{Brain and Mind} \textbf{\bibinfo{volume}{4}},
  \bibinfo{pages}{91} (\bibinfo{year}{2003}).

\bibitem[{\citenamefont{Ascoli and Krichmar}(2000)}]{Ascoli_Krichmar:2000}
\bibinfo{author}{\bibfnamefont{G.~A.} \bibnamefont{Ascoli}} \bibnamefont{and}
  \bibinfo{author}{\bibfnamefont{J.~L.} \bibnamefont{Krichmar}},
  \bibinfo{journal}{Neurocomputing} \textbf{\bibinfo{volume}{48}},
  \bibinfo{pages}{1003:1011} (\bibinfo{year}{2000}).

\bibitem[{\citenamefont{Coelho and da~F.~Costa}(2001)}]{Coelho_Costa:2001}
\bibinfo{author}{\bibfnamefont{R.~C.} \bibnamefont{Coelho}} \bibnamefont{and}
  \bibinfo{author}{\bibfnamefont{L.}~\bibnamefont{da~F.~Costa}},
  \bibinfo{journal}{Neurocomputing} \textbf{\bibinfo{volume}{48}},
  \bibinfo{pages}{555} (\bibinfo{year}{2001}).

\bibitem[{\citenamefont{Waessle
  et~al.}(1981{\natexlab{a}})\citenamefont{Waessle, Peichl, and
  Boycott}}]{Wasslea:1981}
\bibinfo{author}{\bibfnamefont{H.}~\bibnamefont{Waessle}},
  \bibinfo{author}{\bibfnamefont{H.}~\bibnamefont{Peichl}}, \bibnamefont{and}
  \bibinfo{author}{\bibfnamefont{B.~B.} \bibnamefont{Boycott}},
  \bibinfo{journal}{Proc. Roal Soc. London} \textbf{\bibinfo{volume}{212}},
  \bibinfo{pages}{157} (\bibinfo{year}{1981}{\natexlab{a}}).

\bibitem[{\citenamefont{Waessle
  et~al.}(1981{\natexlab{b}})\citenamefont{Waessle, Boycott, and
  Illing}}]{Wassleb:1981}
\bibinfo{author}{\bibfnamefont{H.}~\bibnamefont{Waessle}},
  \bibinfo{author}{\bibfnamefont{B.~B.} \bibnamefont{Boycott}},
  \bibnamefont{and} \bibinfo{author}{\bibfnamefont{R.-B.}
  \bibnamefont{Illing}}, \bibinfo{journal}{Proc. Roal Soc. London}
  \textbf{\bibinfo{volume}{212}}, \bibinfo{pages}{177}
  (\bibinfo{year}{1981}{\natexlab{b}}).

\bibitem[{\citenamefont{da~F.~Costa and Jr}(2001)}]{CostaCesar:2001}
\bibinfo{author}{\bibfnamefont{L.}~\bibnamefont{da~F.~Costa}} \bibnamefont{and}
  \bibinfo{author}{\bibfnamefont{R.~M.~C.} \bibnamefont{Jr}},
  \emph{\bibinfo{title}{Shape Analysis and Classification: Theory and
  Practice}} (\bibinfo{publisher}{CRC Press}, \bibinfo{address}{Boca Raton},
  \bibinfo{year}{2001}).

\bibitem[{\citenamefont{da~F~Costa et~al.}(2002)\citenamefont{da~F~Costa,
  Manoel, Faucereau, Chelly, van Pelt, and Ramakers}}]{LU101}
\bibinfo{author}{\bibfnamefont{L.}~\bibnamefont{da~F~Costa}},
  \bibinfo{author}{\bibfnamefont{E.~T.~M.} \bibnamefont{Manoel}},
  \bibinfo{author}{\bibfnamefont{F.}~\bibnamefont{Faucereau}},
  \bibinfo{author}{\bibfnamefont{J.}~\bibnamefont{Chelly}},
  \bibinfo{author}{\bibfnamefont{J.}~\bibnamefont{van Pelt}}, \bibnamefont{and}
  \bibinfo{author}{\bibfnamefont{G.}~\bibnamefont{Ramakers}},
  \bibinfo{journal}{Network: Comput. Neural Syst.}
  \textbf{\bibinfo{volume}{13}}, \bibinfo{pages}{283} (\bibinfo{year}{2002}).

\bibitem[{\citenamefont{da~F.~Costa and Velte}(1999)}]{Costa_Velte:1999}
\bibinfo{author}{\bibfnamefont{L.}~\bibnamefont{da~F.~Costa}} \bibnamefont{and}
  \bibinfo{author}{\bibfnamefont{T.}~\bibnamefont{Velte}}, \bibinfo{journal}{J.
  Comp. Neurol.} \textbf{\bibinfo{volume}{404}}, \bibinfo{pages}{33}
  (\bibinfo{year}{1999}).

\bibitem[{\citenamefont{da~F.~Costa}(2004)}]{Costa_hier:2004}
\bibinfo{author}{\bibfnamefont{L.}~\bibnamefont{da~F.~Costa}},
  \bibinfo{journal}{Phys. Rev. Lett.} \textbf{\bibinfo{volume}{93}},
  \bibinfo{pages}{98702} (\bibinfo{year}{2004}),
  \bibinfo{note}{cond-mat/0312646}.

\end{thebibliography}

\end{document}